\documentclass[fleqn,usenatbib]{mnras}

\usepackage[T1]{fontenc}

\DeclareRobustCommand{\VAN}[3]{#2}
\let\VANthebibliography\thebibliography
\def\thebibliography{\DeclareRobustCommand{\VAN}[3]{##3}\VANthebibliography}

\usepackage{graphicx}	
\usepackage{amsmath}	
\usepackage{amssymb}	
\usepackage{multirow}

\usepackage{newtxtext,newtxmath}

\usepackage[compact]{titlesec}

\title[LRG satellites]{The satellite population around luminous red galaxies in the 25 $deg^2$ DESI Legacy Imaging Surveys Early Data Release}

\author[M. E. Townsend et al.]{
Melinda Townsend,$^{1}$\thanks{E-mail: mtownsend@ku.edu}
and Gregory Rudnick$^{1}$\thanks{E-mail: grudnick@ku.edu}
\\
$^{1}$University of Kansas, Department of Physics and Astronomy, 1082 Malott,1251 Wescoe Hall Dr., Lawrence, KS 66045 \\
}

\date{Accepted XXX. Received YYY; in original form ZZZ}

\pubyear{2022}

\begin{document}
\label{firstpage}
\pagerange{\pageref{firstpage}--\pageref{lastpage}}
\maketitle

\begin{abstract}
Luminous Red Galaxies, or LRGs, are representative of the most massive galaxies and were originally selected in the Sloan Digital Sky Survey as good tracers of large scale structure. They are dominated by by uniformly old stellar populations, have low star formation rates, early type morphologies, and little cold gas. Despite having old stellar populations and little in situ star formation, studies have shown that they have grown their stellar mass since z=1, implying that they grow predominantly via the accretion of satellites. Tests of this picture have been limited because of the lack of deep imaging data sets that both covers a large enough area of the sky to contain substantial numbers of LRGs and that also is deep enough to detect faint satellites. We 
use the 25 deg$^{2}$ Early Data Release (EDR) of the DESI Legacy Imaging Surveys to characterize the satellite galaxy population of LRGs out to z=0.65. The DESI Legacy Imaging Surveys are comprised of $grz$ imaging to 2-2.5 mag deeper than SDSS and with better image quality. We use a new statistical background technique to identify excess populations of putative satellite galaxies around 1823 LRGs at $0.2<z<0.65$.  In three redshift and luminosity bins we measure the numbers of satellite galaxies and their $r-z$ color distribution down to rest-frame $g$-band luminosity limits at least 3.6 times fainter than $L^*$. In addition, we develop a forward modeling technique and apply it to constrain the mean number of satellites in each of our redshift and luminosity bins.  Finally, we use these estimates to determine the amount of stellar mass growth in LRGs down to the local Universe.
\end{abstract}




\section{Introduction}

The method by which massive galaxies accumulate their mass is an open question in galaxy evolution. Early-type galaxies (ETGs) have low specific star formation rates, old stellar populations, and very little cold gas \citep[e.g.][]{Moustakas2013}. A homogeneous population of early-type galaxies are Luminous Red Galaxies (LRGs). Originally selected as good tracers of large scale structure \citep{Eisenstein2001}, subsequent investigations have shown them to be among the most massive galaxies in the universe, dominated by uniformly old stellar populations \citep{Tojeiro2011}. Despite this, the stellar mass of LRGs has grown by 50\% since z=0.9, implying that they grow mostly through the accretion of passive satellites \citep{Cool2008}. Multiple studies have also shown that the spectra of LRGs indicate that those galaxies have undergone passive fading (e.g. \citep{Cool2008, Banerji2010}). However, some studies find that only bright LRGs evolve passively while fainter LRGs have more extended star formation histories \citep[SFG;][]{Tojeiro2011}.

It is now well accepted that massive galaxies grow hierarchically through the successive effect of many mergers coupled with in situ growth by star formation \citep[e.g.][]{Frenk1991}.  There is overwhelming observational support for this overall picture, with some of the most direct evidence coming from the observations that massive galaxies at $z>0.8$ have  significantly smaller sizes and than similarly massive galaxies in the local universe \citep{Daddi2005,Trujillo2006,VanderWel2008,Buitrago2008,VanderWel2008,VanDokkum2008,Hopkins2009,VanDokkum2010,Newman2012,VanDerWel2014}.  For example, \citet{Trujillo2006} found that massive galaxies at $1.4<z<1.7$ were at least a factor of four times smaller in the rest-frame $V$-band than local counterparts of the same stellar mass, and that the the stellar density of these objects are at least 60 times larger than present-day massive ellipticals, indicating that they must have grown predominantly through dry mergers and \citet{Daddi2005} hypothesized that the size growth came from the accretion of satellite galaxies.

Likewise, \citet{Man2016} found that, to explain the observed number density evolution of massive galaxies, minor mergers are a necessary component to bring compact quiescent ellipticals into agreement with the stellar mass-size relation. Accretion-based growth of passive galaxies leads to "inside-out" growth, where massive galaxies grow via a series of minor mergers and a build-up of extended stellar halos \citep{VanDokkum2010}. Using luminosity functions of LRGs and their satellites, \citet{Tal2012a} determined that the mass ratio of LRGs to satellites is 4:1, making mass growth through major mergers unlikley. This is consistent with the findings in \citet{Bezanson2009}, which find that central stellar densities of high redshift ellipticals are only a factor of a few higher when compared to local ellipticals even though the total stellar density of high redshift ellipticals are orders of magnitude higher than their local counterparts, indicating that mass is being deposited in the outskirts of the galaxy through minor mergers.  \citet{VanDeSande2013} found that for fixed dynamical mass of massive quiescent galaxies from z $\sim$ 2, the mass density within one effective radius decreases by a factor of 20 while within a fixed physical radius of 1 kpc the mass density decreases by a factor of $\sim$ 2, which is consistent with inside-out growth through minor mergers.

These observational results are supported by simulations of galaxy growth.  \citet{Naab2010}, using a high-resolution hydrodynamical cosmological simulation, shows that the accretion of weakly bound material - or minor mergers - can cause the radius of a massive spheroidal galaxy to increase as the square of the mass, whereas major mergers would cause the radius to increase at a linear rate. This result echos the observations and implies a cause; namely, that early-type galaxies are more compact at earlier times and grow less compact at low redshift because of the accretion of low-mass satellites.

The rest-frame $B$-band luminsosity function of red galaxies also points to mergers as a mechanism to build up mass in massive systems. \citet{Bell2004} presented the rest-frame colors and luminosities of $\sim$ 25,000 galaxies in 0.2 $<$ z $<$ 1.1 from the 0.78 deg$^{2}$ Classifying Objects by Medium-Band Observations in 17 Filters (COMBO-17) survey. They found that the $B$-band luminosity density does not evolve significantly in this redshift range, which, when coupled with the passive fading of the galaxy stellar populations, implies that there has been a build-up of stellar mass in the non-star forming population by a factor of 2 since z $\sim$ 1. \citet{Faber2007} came to a similar result in their study that compares the luminosity functions from \citet{Willmer2006} of red and blue galaxies out to z $\sim$ 1 of the DEEP2 and COMBO-17 surveys. They also found a virtually constant $B$-band luminosity density for red galaxies since z $\sim$ 1, while the luminosity density of blue galaxies falls by $\sim$0.6 dex. They argue that dry mergers are involved in building up present day massive ellipticals. Similarly, \citet{Brown2007} used the luminosity density and number density of galaxies to find that the amount of stellar mass contained in $L*$ red galaxies has doubled since z $=$ 1, but that the stellar mass in 4$L*$ red galaxies evolved more slowly.

In contrast, \citet{DePropris2010} derived an upper limit for a dry merger rate in their measurement of the fraction of LRGs at 0.45 $<$ z $<$ 0.65 in dynamically close pairs taken from the 2dF-SDSS LRG and QSO (2SLAQ) redshift survey. They find that minor dry mergers (a luminosity ratio 1:4 or higher) are unimportant to the mass build-up of the red sequence at z $<$ 0.7. This holds at higher redshift, as well; mergers are found to not be the dominant channel for stellar mass build-up in early-type galaxies out to z $<$ 0.8 \citep{Cimatti2006} and for quiescent galaxies out to z $<$ 1 \citep{Moustakas2013}. \citet{Scarlata2007} determined that dry mergers are likely not a significant contributor to the build-up of massive ETGs, and that the most massive of these galaxies were already assembled 8 Gyr ago (although fainter ETGs keep assembling mass from z = 0.7 to the present \citep{Tojeiro2011}. \citet{Banerji2010} finds that the stellar mass function for LRGs with  $M_\star> 3 \times 10^{11} {\rm M}_{\odot}$ shows little evolution between 0.4 $<$ z $<$ 0.8, suggesting that most massive systems were in place by z $=$ 0.8 \citep[see also][]{Huertas-Company2016}.  

There are theoretical reasons to believe that understanding accretion-based growth of massive ellipticals are important to understanding how galaxies evolve. \citet{DeLucia2007} found that of z = 0 brightest cluster galaxies (BCGs) in the Millennium Simulation, only 10 percent were formed before z $\sim$ 1, and half were assembled after z $\sim$ 0.5. They also find that 50 percent of the stars found in BCGs were already formed by z $\sim$ 5. Old stellar populations coupled with the late assembly times of these galaxies indicate that the BCGs in their sample gained most of their mass through accretion of satellite galaxies. Using mock catalogs constructed from the halo occupation distribution framework and comparing to data from the Bo\"{o}tes field for galaxies between 0.5 $<$ z $<$ 0.9, \citet{White2007} found that their models overpredict the number of satellites that populate massive halos and interpret this as evidence that massive satellite galaxies are merging or otherwise being disrupted by the central galaxy in that redshift range. Building off this study, \citet{Brown2008} measures the halo occupation distribution for red galaxies in the Bo\"{o}tes field and found that, while most massive galaxy stellar mass growth occurs prior to z = 1, massive galaxy growth continues and that a typical central galaxy grows by 30 percent from z < 1.

Although there is not uniform agreement, there are many indications in both observations and simulations that merger events are responsible for the mass build up of massive ellipticals, but how this growth depends on redshift and primary galaxy mass is inconclusive. In addition, it is unclear what role is played by faint satellites. Most previous studies only probe satellites that are relatively massive. \citet{DePropris2010} and \citet{Bell2006} focused on the mass build up of early-type galaxies through major mergers. \citet{Bundy2009} probed satellites out to z $\sim$ 1.4 to log(M$_{*}$/M$_{\odot}$) = 9.8, but for a combined field of less than a degree.

It has been difficult to study faint satellites because of the dearth of data that can identify those faint satellites while at the same time encompassing a large sample of giant ellipticals. This situation is changing with the arrival of the DESI Legacy Imaging Surveys (Legacy Survey) \citep{Dey2018}. While large surveys have revolutionized the field of galaxy evolution, there has not until recently been data sets that both cover a large enough area of the sky to contain a substantial number of luminous ellipticals and deep enough to detect faint satellites out to intermediate redshifts. Surveys like SDSS are too shallow to see very faint objects around luminous galaxies, and surveys like the NOAO Deep Wide-Field Survey only cover a small fraction of the sky. In this paper, we combine the deep photometry of the Legacy Survey and the spectroscopic data from SDSS to characterize the satellite population around SDSS-identified LRGs 2 to 2.5 magnitudes deeper than SDSS. By combining these two surveys, we solve the problem of shallow survey depth and limited sky coverage. We use the two surveys in the following way: We use SDSS spectroscopy to select LRGs and the Legacy Survey imaging to detect faint candidate satellite galaxies. We use statistical background techniques to isolate likely satellites and study the abundance and properties of those satellites. 

This paper is structured as follows: Sec. \ref{sec:data} describes the data and sample selection, Sec. \ref{sec:analysis} describes our analysis method, we present our results in Sec. \ref{sec:results}, and discuss these results in Sec. \ref{sec:discuss}. In Sec. \ref{sec:concl} we present our summary and conclusions. We assume H$_{0}$ $=$ 69.6, $\Omega$$_{m}$ $=$ 0.286, and $\Omega$$_{\Lambda}$ $=$ 0.714 \cite{Bennett2014}).

\section{Data Description}
\label{sec:data}

\begin{figure}
	\includegraphics[width=\columnwidth]{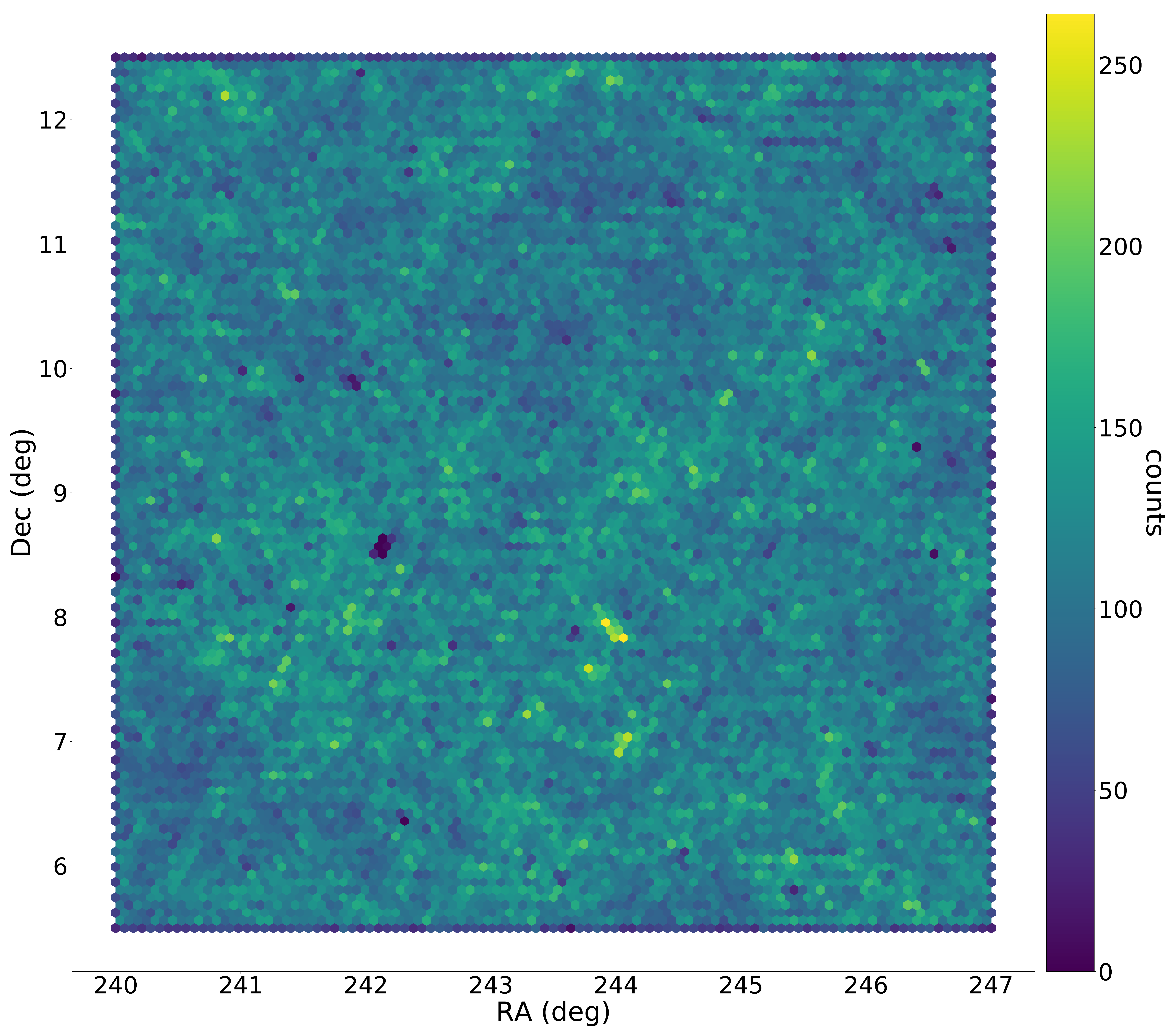}
    \caption{Density maps of the EDR after we eliminate Legacy Survey sources with observed $z$-band magnitude $\geq$ 22.75. The density variations reflect true variations in the large scale structure of galaxies.}
    \label{fig:LS-density}
\end{figure}

In this study we use SDSS-identified Luminous Red Galaxies using spectroscopic redshifts of SDSS and photometry from the DESI Legacy Imaging Surveys.

\subsection{SDSS Luminous Red Galaxies}
\label{sec:sdss} 

The sample of LRGs were identified in the Baryon Oscillation Spectroscopic Survey (BOSS), part of the SDSS-III project. The BOSS LRGs are divided into a low redshift sample (LOWZ) and a constant-mass sample (CMASS) \citep{Reid2016}. Both the LOWZ and CMASS samples were selected based on a set of color-magnitude cuts. For the LOWZ sample, these cuts are designed to only select the brightest and reddest galaxies at low redshift (z $<$ 0.4) and are similar to the SDSS- I/II Cut-I Luminous Red Galaxies. The LOWZ sample is also approximately volume-limited over 0.2 $<$ z $<$ 0.4 and has a constant space density of $\sim$3 $\times$ 10$^{-4}$ h$^{3}$Mpc$^{-3}$. The CMASS sample includes galaxies between 0.4 $<$ z $<$ 0.7 with an approximately constant stellar mass limit over the redshift range. The CMASS sample is also selected by a color cut, similar to SDSS-I/II Cut-II and 2SLAQ LRGs. However, the cuts are bluer and more faint to increase the number density of targets in the CMASS redshift range. Higher redshift galaxies in CMASS are isolated using ($g-r$) and ($r-i$) colors. Together, LOWZ and CMASS make a spectroscopic sample that is 80 percent complete at log$_{10}$(M/M$_{\odot}$) $\geq$ 11.6 at z $<$ 0.61. We choose to study LRGs out to 0.65. There are 151 LRGs between 0.61 $<$ z $<$ 0.65 and may bias our sample in favor of older LRGs. However, we ran our analysis on samples with the maximum redshift of 0.65 and 0.61 and found no difference in our results. For more details on the selection criteria for LOWZ and CMASS (see \cite{Reid2016}). Spectroscopic redshifts for this study are from SDSS DR14.

We identify 1,823 LRGs in the EDR in the redshift range 0.2 $<$ z $<$ 0.65. We tested to see if LRG-LRG pairs skew our analysis. We explicitly removed all LRGs that had an LRG within the approximate virial radius in redshift slices of 0.003. Of the 1,823 LRGs in this sample, 92 were removed because they were part of an LRG-LRG pair, leaving us with a sample of 1,731 LRGs. The original 1,823 is our ``pair'' sample, because it includes LRG-LRG pairs, and the 1,731 LRGs is our ``no pair'' sample. We calculated the distribution of satellite galaxies for both the ``pair'' and ``no pair'' sample and found that they were consistent with each other to within one standard deviation. We conclude, then, that including LRG-LRG pairs does not alter our results.

\begin{figure*}
	\includegraphics[width=0.8\textwidth]{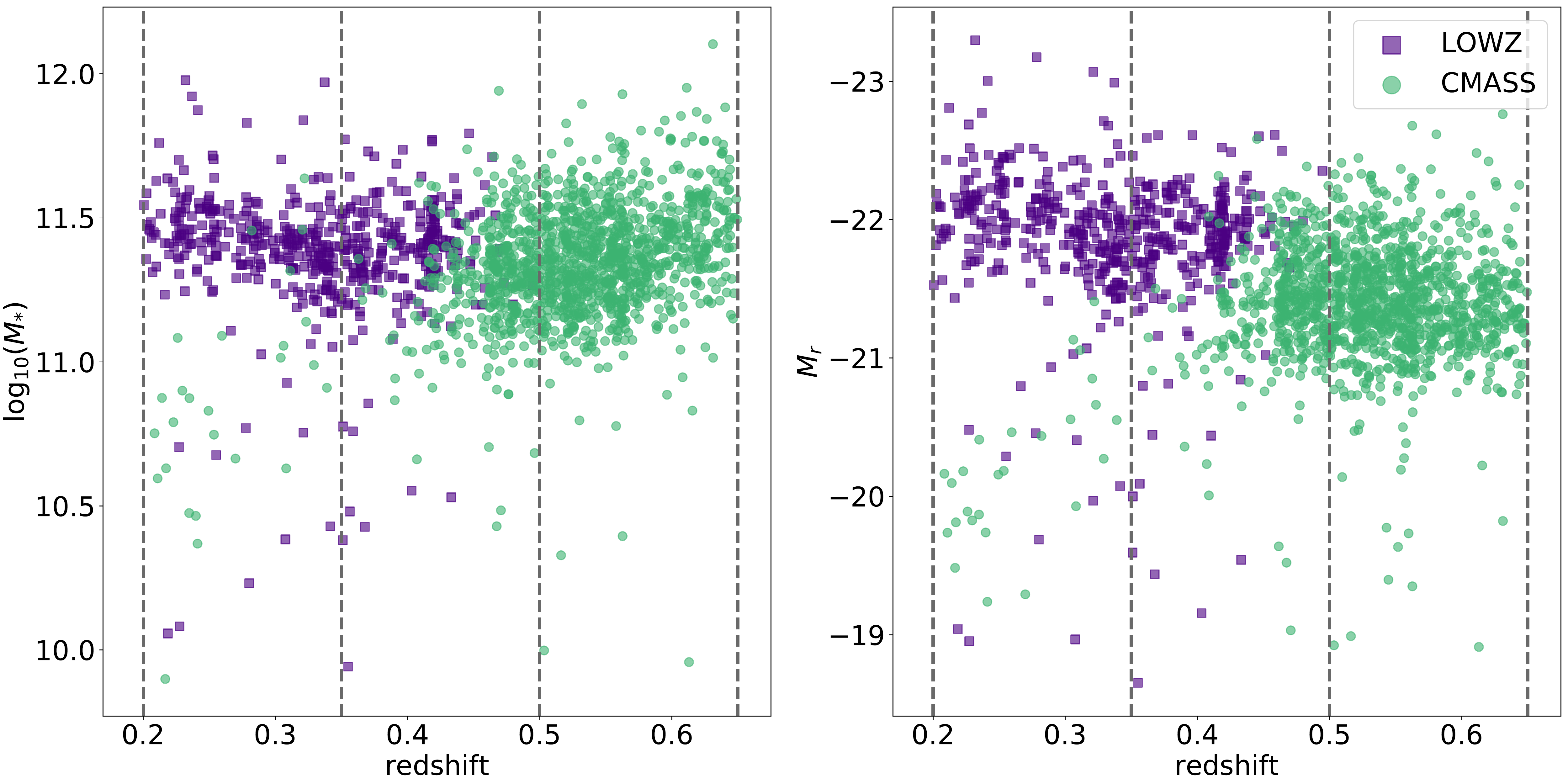}
    \caption{The plot shows the log$_{10}$(M$_{\star}$) and absolute $r$-band magnitude of LRGs in our sample v. LRG redshift in the left- and right-hand panels, respectively. LOWZ LRGs are indicated by indigo squares and CMASS LRGs are indicated by green circles. Vertical lines show the bounds of our redshift bins.}
    \label{fig:LRG_info}
\end{figure*}

\subsection{Legacy Survey Description}
\label{sec:legacy}

The DESI Legacy Imaging Surveys\footnote{https://www.legacysurvey.org/} are a trio of surveys that image $\approx$  14,000 deg$^{2}$ of the extragalactic sky visible from the northern hemisphere in three optical bands - $g$, $r$, and $z$ - with a combined survey footprint that is split into two contiguous regions by the galactic plane \citep{Dey2018}. Those surveys are the Dark Energy Camera Legacy Survey, the Beijing-Arizona Sky Survey, and the Mayall $z$-band Legacy Survey. The data used in this paper comes from the Dark Energy Camera Legacy Survey.

The Dark Energy Camera Legacy Survey (DECaLS) utilizes the Blanco 4-m telescope at Cerro Tololo Inter-American Observatory using the Dark Energy Camera and covers 9,000 deg$^2$ in both the Northern and Southern Galactic Cap. The survey obtains optical imaging data in the $g$-, $r$-, and $z$-bands, overlaps existing spectroscopy from the Sloan Digital Sky Survey (SDSS), and reaches 5$\sigma$ point source depths of 24.0, 23.4, and 22.5 in the $g$-, $r$-, and $z$-band, respectively, more than two magnitudes deeper than SDSS. Source detection is performed using The Tractor\footnote{http://thetractor.org/}, a forward modeling algorithm in which each source is modeled at the pixel level through simultaneous fits to a set of individual images. Each source is modeled using a set of parametric light profiles: a delta function, deVaucouleurs r$^{-1/4}$ law, exponential disk, or exponential disk plus deVaucouleurs. See \cite{Dey2018} for a more detailed explanation. For this study we use the photometry from the 8th data release\footnote{https://www.legacysurvey.org/dr8/}.


For this study we use data from the area covered by the Legacy Survey Early Data Release (EDR), which covers between 240 and 245 degrees in right ascension and between 6.5 and 11.5 degrees in declination. Fig. \ref{fig:LS-density} illustrates the z-band source density in the field of study. We want to establish a uniform detection limit across the survey area to make sure we are equally sensitive to satellites for all LRGs. To do this, we use the 5$\sigma$ galaxy detection limit in the z-band to find the 90 percent brightest galaxy depth, and we limit our sample to galaxies brighter than this limit. We find this limit by binning sources into equal-size pixels using HEALPix\footnote{http://healpix.sourceforge.net} \citep{Gorski2005, Zonca2019} and finding the median depth in each pixel. HEALPix discretizes the surface of a sphere into equal-area, non-overlapping tiles called pixels and astrophysical analysis can be done on a per-pixel basis. Then we determine the 10th percentile and use that as our magnitude limit. We find this limit to be a $z$-band magnitude 22.75 mag. We also confirm that all sources detected in the $z$-band are also detected in the $g$- and $r$-band. The results of this $z$-band magnitude cut are shown in Fig. \ref{fig:LS-density}. The density map for our magnitude complete sample demonstrates the presence of true large scale structure variations in our field. As we will describe in Sec. \ref{sec:stat-sub}, these variations motivate our use of a local background subtraction technique.

\subsection{Rest Frame Magnitudes and Masses for LRGs}
\label{mags-and-masses}

To further characterize the SDSS LRGs, we compute both the rest frame $r$-band magnitudes and the stellar masses. We compute the rest-frame magnitudes for BOSS LRGs using the photometric redshift code EAZY \citep{Brammer2008}, fixing each LRG at its spectroscopic redshift and using as inputs the $grz$ photometry from the Legacy Surveys, as well as WISE bands $W1$, $W2$, $W3$, and $W4$. LRG stellar masses are computed using iSEDfit from \cite{Moustakas2013} assuming a Kroupa IMF \citep{Kroupa2001}. iSEDfit uses Baysian inference to extract physical properties from a galaxy’s observed broadband spectral energy distribution. 

Fig. \ref{fig:LRG_info} shows the stellar mass and r-band absolute magnitude as a function of redshift for SDSS LRGs. The stellar mass (left plot) remains constant with redshift, although there is some scatter into lower stellar mass. Absolute $r$-band magnitude also shows scatter at low redshift toward fainter magnitudes. We have 516 LRGs from the LOWZ sample and 1,307 from CMASS, for a total sample of 1,823 LRGs. Of these, 3.7 percent have stellar masses below 10$^{11}$ with 19 appearing in the LOWZ sample and 49 appearing in the CMASS sample, and 3.3 percent have M$_{r}$ $>$ -20.5 with 19 appearing in the LOWZ sample and 41 appearing in the CMASS sample. Visual inspection of these outliers show that these galaxies are predominantly red elliptical galaxies.

\subsection{Luminosity Completeness}
\label{sec:lum-completeness}

The redshift range of our LRG sample is 0.2 $<$ z $<$ 0.65. We wish to be equally complete to satellites above a given luminosity limit over a range in redshift.  We decide to split our sample into three luminosity-complete subsamples, each spanning a different range in redshift.  We determine the corresponding luminosity limit for each subsample using the UltraVISTA catalog from \cite{Muzzin2013}, which is an ultra-deep $K_s$-selected catalog in the COSMOS field. The catalog covers 1.62 deg$^{2}$ and has photometry from 30 bands, to much deeper limits than the Legacy Surveys. We use this catalog to determine the 90 percent completeness limit in the rest-frame $g$-band luminosity for three redshift bins: 0.2 $<$ z $<$ 0.35, 0.35 $<$ z $<$ 0.5, and 0.5 $<$ z $<$ 0.65. As we wish to determine our luminosity in the DECam $g$-band filter, we use EAZY to synthesize $g$-band rest-frame luminosities using the DECam filter curve \citep{Brammer2008}. We note that we do not use the $r$-band because our satellites are too faint to be detected in WISE $W1$ and $W2$ and therefore the $r$-band photometry would be unconstrained. LRGs are bright enough to be detected in WISE, so we use the $r$-band luminosity for them. 

Using only the UltraVISTA galaxies within our redshift range, we measure the luminosity down to which we recover 90 percent of the sources brighter than that limit at the high end of each redshift bin, subject to our observed $z$-band magnitude limit of $\leq$ 22.75. We go through this process for each redshift bin and for all UltraVISTA sources, red UltraVISTA sources, and blue UltraVISTA sources (splitting the red and blue populations at $(U-V)=1.4$). We chose the most conservative luminosity where 90 percent of the galaxies were recovered for each redshift bin.

\begin{figure}
	\includegraphics[width=\columnwidth]{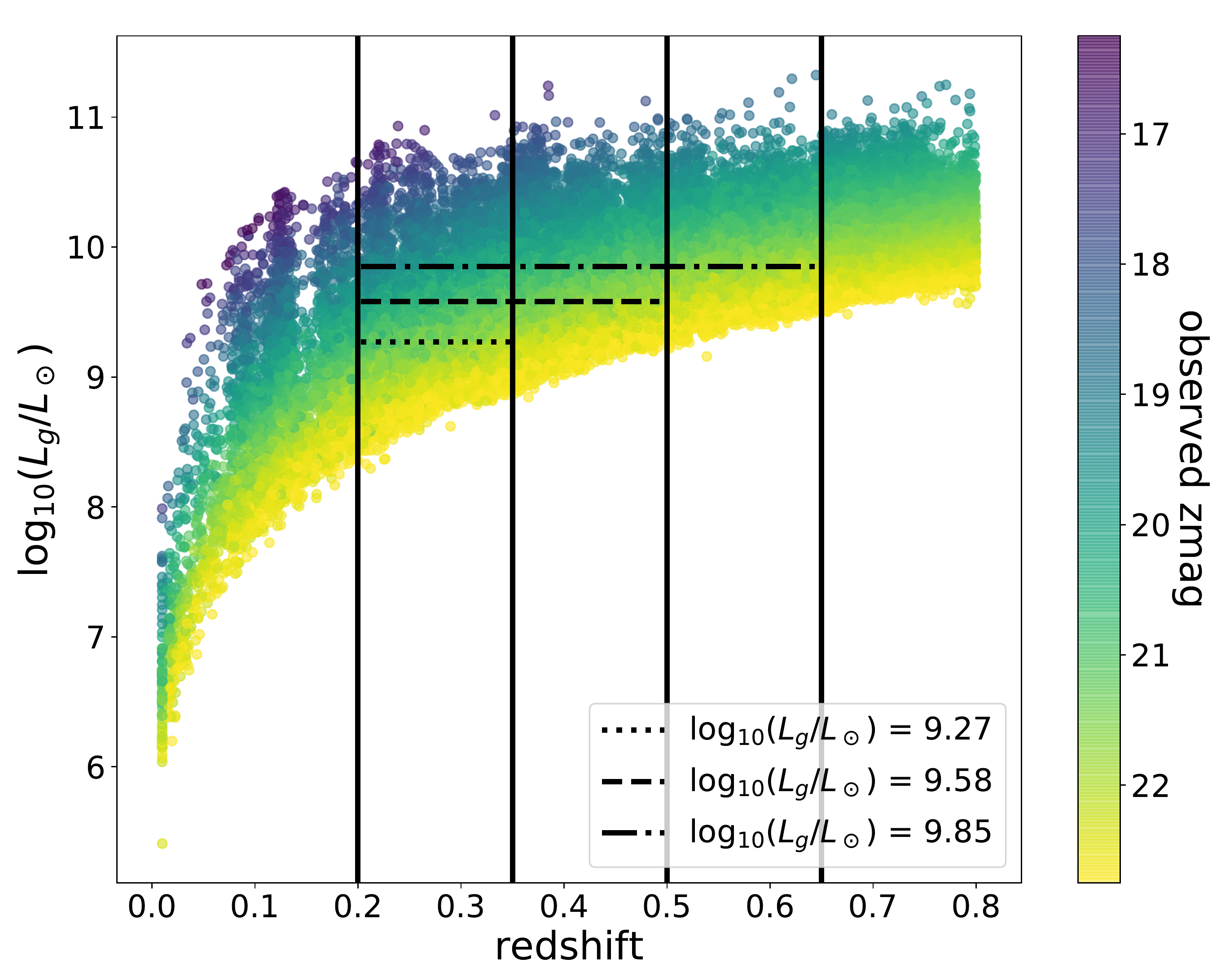}
    \caption{The log$_{10}$($L$$_{g}$/$L$$_{\odot}$) of the UltraVISTA galaxies at z $<$ 1 and with an observed $z$-band magnitude brighter than 22.75. The solid lines denote our redshift bins. Each horizontal line marks the 90 percent luminosity completeness for different redshift ranges. The colorbar indicates the observed z-band magnitude of the sources. This defines three luminosity complete subsamples: for 0.2 $<$ z $<$ 0.35 the luminosity limit is log$_{10}$($L$$_{g}$/$L$$_{\odot}$) $\geq$ 9.27, for 0.2 $<$ z $<$ 0.5 the luminosity limit is log$_{10}$($L$$_{g}$/$L$$_{\odot}$) $\geq$ 9.58, and for 0.2 $<$ z $<$ 0.65 the luminosity limit is log$_{10}$($L$$_{g}$/$L$$_{\odot}$) $\geq$ 9.85.}
    \label{fig:UV-lum}
\end{figure}

Fig. \ref{fig:UV-lum} shows the log$_{10}$($L$$_{g}$/$L$$_{\odot}$) of the $z$-band magnitude $\leq$ 22.75 UltraVISTA galaxies out to z $=$ 1, with the solid vertical lines deliniating the boundaries of our redshift bins. The three horizontal lines indicate the luminosity limit for each redshift bin and the colorbar shows the observed $z$-band magnitude of the UltraVISTA sources. We have three luminosity complete samples for three redshift bins: For 0.2 $<$ z $<$ 0.35 the luminosity limit is log$_{10}$($L$$_{g}$/$L$$_{\odot}$) $\geq$ 9.27, for 0.2 $<$ z $<$ 0.5 the luminosity limit is log$_{10}(L_{g}/L_{\odot}) \geq$ 9.58, and for 0.2 $<$ z $<$ 0.65 the luminosity limit is log$_{10}$($L$$_{g}$/$L$$_{\odot}$) $\geq$ 9.85. Notice that the depth of the Legacy Surveys make it possible to detect faint galaxies all the way down to log$_{10}$($L$$_{g}$/$L$$_{\odot}$) $=$ 9.85 even in the upper redshift bin. In \citet{Rudnick2009}, they construct luminosity functions for red sequence galaxies from 0.4 $<$ z $<$ 0.8. At $z \sim 0.6$, they found that M$^\star_{g} \sim -21$ which corresponds to log$_{10}$($L$$_{g}$/$L$$_{\odot}$) $=$ 10.43, or 3.6 times higher than our limit here in our highest redshift bin. This means at our high redshift limit we are 3.6 times lower than L*.

\section{Determination of the number of satellite galaxies}
\label{sec:analysis}

The Legacy Survey $grz$ photometry is not sufficient to achieve precise photometric redshfits for our candidate satellite galaxies and obtaining complete spectroscopy to these depths and over even just the EDR would be impractical. Therefore we quantify the satellite population around LRGs by implementing a statistical background subtraction method. We describe this method in the following subsections.

\subsection{Statistical Background Subtraction}
\label{sec:stat-sub}

The foundation of our statistical background subtraction method consists of three parts: counting the number of near neighbors within a defined search radius, determining how many background sources to expect, and subtracting the second number from the first for each LRG.

We set our search radius to correspond to the expected virial radius ($R_200$) for our LRGs.  To estimate this radius we used the stellar mass to halo mass relation and estimated the virial radius from the halo mass. The 75th and 25th percentile of LRGs masses are separated by 0.2 dex and the median mass does not change with redshift, so we assume that LRGs have a fixed stellar mass of log$_{10}$(M/M$_{\odot}$) $=$ 11.3. We use equations 2 and 11-14 in \cite{Moster2013}, which are reproduced below, where we have slightly changed the original notation for clarity:

\begin{equation}
    \frac{M_\star}{M_{h}} = 2N \left[  \left( \frac{M_h}{M_{1}}^{\beta} \right) + \left(\frac{M_h}{M_{1}}^{\gamma} \right) \right]^{-1}
\end{equation}
\begin{equation}
    {\rm log}~M_1(z) = M_{10} + M_{11}\frac{z}{z+1}
\end{equation}
\begin{equation}
    N(z) = N_{10} + N_{11}\frac{z}{z+1}
\end{equation}
\begin{equation}
    \beta(z) = \beta_{10} + \beta_{11}\frac{z}{z+1}
\end{equation}
\begin{equation}
    \gamma(z) = \gamma_{10} + \gamma_{11}\frac{z}{z+1}
\end{equation}
where $M_\star$ and $M_h$ are the stellar and halo mass respectively, $M_{1}$ is the characteristic mass, $N$ is the normalization, and $\beta$ and $\gamma$ are the power law slopes. We use best fit values for $M_{10}$, $M_{11}$, $N_{10}$, $N_{11}$, $\beta_{10}$, $\beta_{11}$, $\gamma_{10}$, and $\gamma_{11}$ from Table 1 in \cite{Moster2013}. Using this relation we find log($M_h/M_\odot)=13.93$, 13.99, and 14.04 in our three redshift bins.  We therefore adopt log($M_h/M_\odot)\approx 14.0$ for all LRGs.  Using this, we estimate $R_{200}$ using the critical density ($\rho_c$) at each epoch and the following equation:
\begin{equation}
R_{200} = \left(\frac{4\pi(200\rho_c)}{3M_h}\right)^{1/3}.
\end{equation}
The corresponding values of R$_{200}$ for our sample is are 0.57, 0.59, and 0.64~Mpc in our three redshift bins.  We therefore assume that the virial radii for our LRGs can be approximated by $R_{200}\approx 0.6$~Mpc and we count satellites within this radius.

To count the number of near neighbors N$_{nn}$ around each LRG, we used the k-d tree algorithm from scikit-learn \citep{scikit-learn} to find sources projected within a 600 kpc radius, converted to a unique angular distance for each LRG according to \citet{2006PASP..118.1711W}. 

To estimate the background and foreground contamination, we create a HEALPix \citep{Gorski2005, Zonca2019} map of all galaxies in the EDR using pixel dimensions of 0.00082 deg$^{2}$ and define an annulus around each LRG between 0.4 and 0.5 degree radius. We determine that, at this distance, deviations from large scale structure start to dominate over the Poisson uncertainties. Because a fixed angular scale will measure different background at different redshifts, we test multiple apertures and find that our results are insensitive to background aperture size. We identify the HEALPix pixels within the background annulus and use the galaxies within those pixels as our background sources.  This is advantageous as HEALPix allows us to very quickly index which galaxies fall in which part of the sky. We do this separately for each LRG, which allows our background estimate to trace the local large scale structure on which each LRG lies.  At the median redshift of each of our redshift bins, 0.45 degrees, corresponds to 9.96 Mpc, well beyond the size of the largest virialized clusters. 

\begin{figure}
	\includegraphics[width=\columnwidth]{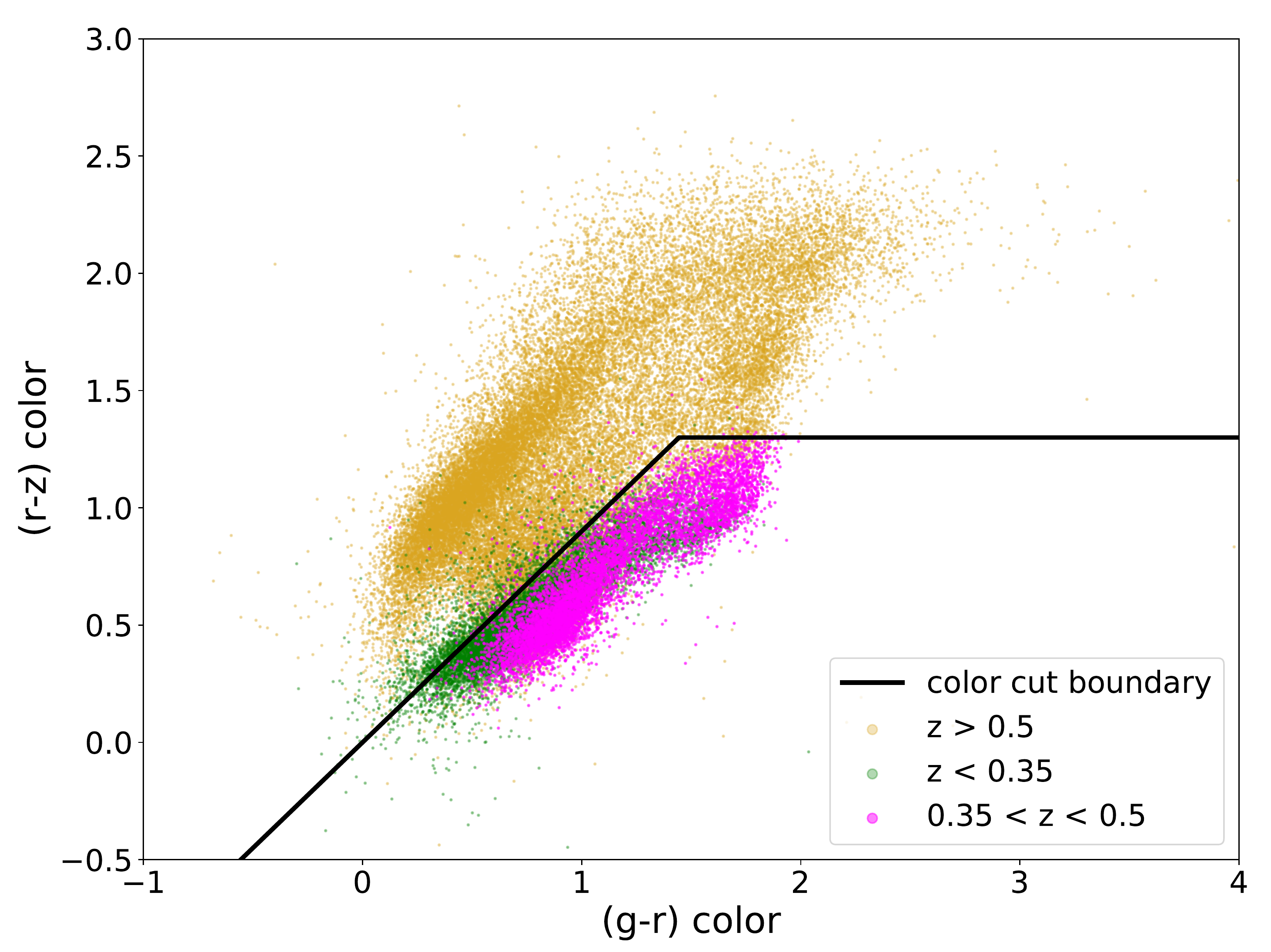}
    \caption{An example of the color-color cuts made for redshift bin 0.35 $<$ z $<$ 0.5 using galaxies from the UltraVISTA catalog. The cuts primarily eliminate galaxies that are at higher redshift than the target redshift slice.}
    \label{fig:cc-cut-ex}
\end{figure}

To minimize background and foreground contamination, we use the UltraVISTA catalog make cuts in observed ($g-r$) and ($r-z$) color for each redshift bin. We make cuts that maximize the number of galaxies in our redshift range and minimize the galaxies outside our redshift range. For example, we find that galaxies with ($r-z$) $<$ 0.9($g-r$) and ($r-z$) $<$ 1.3 are in the redshift range 0.35 $<$ z $<$ 0.5. We reduce background and foreground contamination by only considering galaxies within these bounds at these redshifts. For redshift bins 0.2 $<$ z $<$ 0.35 the cuts are made at ($r-z$) $=$ ($g-r$) and ($r-z$) $=$ 1.1 and for redshift bin 0.5 $<$ z $<$ 0.65 the cuts are made at ($r-z$) $=$ ($g-r$) + 0.2 and ($r-z$) $=$ 1.6. An example of these cuts can be seen in Fig. \ref{fig:cc-cut-ex}. Table \ref{tab:cc-cut-stats} reports the percentage of target sources retained by the cut, as well as the percentage of sources retained that are actually outside the redshift slice. This method was most effective at eliminating sources above our target redshift range. These cuts are implemented at all stages of the analysis, and significantly reduce the Poisson uncertainty in our background estimation.

\begin{table*}
	\centering
	\caption{Results of color-color cut for each redshift bin. The table reports the percentage of sources that are truly in the redshift bin that are retained by the cut, as well as the percentage of interloper sources retained that are actually outside the redshift slice. This method was most effective at eliminating sources above our target redshift range.}
	\label{tab:cc-cut-stats}
	\begin{tabular}{cccc} 
		\hline
		\multicolumn{4}{c|}{Results of color-color cut} \\
		\hline
       redshift bin & percentage of true sources retained  & percentage of low-z interlopers retained & percentage of high-z interlopers retained \\
		\hline
		0.2 $<$ z $<$ 0.35 & 96.6\% & 81.6\% & 27.3\%\\
		0.35 $<$ z $<$ 0.5 & 97.4\% & 91.8\% & 16.6\%\\
		0.5 $<$ z $<$ 0.65 & 96.7\% & 98.8\% & 11\%\\
		\hline
	\end{tabular}
\end{table*}

For each LRG, we place near neighbors, background, and foreground galaxies in their own color-color-magnitude diagram, using their observed ($r-z$) and ($g-r$) colors and observed $z$-band magnitude. The grid consists of 50x50x50 cells, spanning -1.8 $<$ $(r-z)$ $<$ 10.4, -6.5 $<$ $(g-r)$ $<$ 10.6, and 13 $<$ $z$-band magnitude $<$ 23, corresponding to the range of our values in our catalog. The number of background galaxies N$_{bkg}$ in each color-color-magnitude cell is then scaled to the angular area of the 600 kpc search radius to reflect the number of background galaxies we would expect to see within that area.

In performing this normalization, we account for the amount of area of the sky lost due to bright stars and large foreground galaxies in two ways. For the background, we identify Legacy Survey galaxies that are flagged for touching a Tycho-2 or Gaia star (to a $g$-band magnitude $<$ 13), a large galaxy, or a star cluster. We determine in which HEALPix pixels in the background annulus the flagged sources are found and exclude the flagged pixel from our analysis. The median percentage of our background area with compromised photometry is less than two percent.

For near neighbors, we use a different method because the area of our HEALPix pixels is too large and results in the elimination of too much area within the 600 kpc search radius. Furthermore, we find that the area close to our LRGs are not contaminated by bright foreground galaxies or star clusters, so the only contamination is from bright stars. Rather than using HEALPix, we use the k-d tree algorithm to find Gaia DR2 stars within our 600 kpc search radius. When a star is found within the search radius, the region affected by the star's halo is calculated by Equation \ref{gaia-radius} as prescribed by the Legacy Survey pipeline\footnote{https://github.com/legacysurvey/legacypipe}

\begin{equation}
    \label{gaia-radius}
    R_{G} = 150 \times 2.5^{11 - gmag_{G}} \times (0.262/3600)
\end{equation}

where R$_{G}$ is the resulting radius in degree and gmag$_{G}$ is the observed $g$-band magnitude for the star in the Gaia DR2 catalog. We then use the radius to determine the area lost to contamination and remove any flagged near neighbor sources from the analysis. The median percentage of the search areas lost to contamination is less than nine percent. Both of these contamination calculations are done for each LRG. Near neighbors and the background are then scaled by these area modifications. 

Once these area corrections are made, we calculate the number of satellites by subtracting the number of background galaxies from the number of near neighbor galaxies, N$_{sat,i}$  = N$_{nn,i}$ - N$_{bkg,i}$, on a cell-by-cell basis in the color-color-magnitude diagram. We sum over all color-color-magnitude bins to get a total number of satellite galaxies for each system. Formally, this number can be negative due to the variation of the density of the local background. However, these measurements can be used to get a statistical estimate of the number of satellite galaxies around LRGs at different redshifts. To measure satellite numbers for our luminosity complete samples, we divide the LRGs into three redshift bins: 0.2 $<$ z $<$ 0.35, 0.35 $<$ z $<$ 0.5, and 0.5 $<$ z $<$ 0.65. We then apply selection matrices to each resulting color-color-magnitude diagram to determine the number of satellites per LRG above the corresponding luminosity limit. Selection matrices are described in the next subsection.

\subsection{Satellite Identification in Bins of Luminosity and Redshift}
\label{sec:masks}

In Sec. \ref{sec:lum-completeness} we describe our luminosity completeness limits, derived from the UltraVISTA catalog. One way to measure the rest-frame luminosity of our satellite galaxies would be to assume that all galaxies are at the LRG redshift and to then fit the sparsely sampled SEDs to derive rest-frame luminosities. However, the vast majority of galaxies in close projection around any LRG are not at the LRG redshift and this would cause significant systematic errors in the luminosity \citep{Rudnick2009}. Instead, we decide to establish a color-color dependent $z$-band magnitude limit, brighter than which galaxies at the LRG redshift would be above our luminosity limit. We map these luminosity completeness limits onto an observed $z$-band magnitude. We perform this translation of rest-frame to observed properties, as our lack of redshift information for satellite galaxies makes it impossible to directly compute their rest-frame luminosities. To do this, we first match the Legacy Survey catalog to UltraVISTA so we can use Legacy Survey photometry and UltraVISTA redshifts. Then, in each ($r-z$) and ($g-r$) cell we determine the median observed $z$-band magnitude of UltraVISTA galaxies that are within $\Delta~{\rm log}_{10}(L_{g}/L_{\odot}) = 0.2$ of the luminosity limit and in bins of 0.05 in redshift in each color-color cell. We do this for each combination of luminosity limit and redshift. This median $z$-band magnitude in each color-color cell defines a selection boundary in color-color space brighter than which galaxies at the redshift of the LRG are luminosity complete. For example, if a color-color cell has a selection limit of $z=20$ at the redshift of the LRG, then only cells brighter than that limit will contribute to the total satellite count. This method assumes that all galaxies around the LRG that are in excess above the background are at the redshift of the LRG. The color dependence of the $z$-band magnitude limit accounts for the SED shape variations among satellite galaxies.

\section{Results}
\label{sec:results}

\subsection{\texorpdfstring{Distribution of Measured $N_{\textbf{sat}}$}{Distribution of Measured \textbf{Nsat}}}
\label{sec:nsat}

We find the number distribution of satellite galaxies N$_{sat}$ in each redshift and luminosity bin. For LRGs at redshift 0.2 $<$ z $<$ 0.35 we find the distribution for luminosity bins log$_{10}$($L$$_{g}$/$L$$_{\odot}$) $\geq$ 9.27, log$_{10}$($L$$_{g}$/$L$$_{\odot}$) $\geq$ 9.58, and log$_{10}$($L$$_{g}$/$L$$_{\odot}$) $\geq$ 9.85. For LRGs in the redshift range of 0.35 $<$ z $<$ 0.5 we show the distribution for the log$_{10}$($L$$_{g}$/$L$$_{\odot}$) $\geq$ 9.58 and log$_{10}$($L$$_{g}$/$L$$_{\odot}$) $\geq$ 9.85 bins. For LRGs in the redshift range 0.5 $<$ z $<$ 0.65 we show the distribution for the log$_{10}$($L$$_{g}$/$L$$_{\odot}$) $\geq$ 9.85 luminosity bin, only. These distributions can be found in the appendix, and they represent the number distribution of possible satellite galaxies above a certain luminosity threshold for LRGs in the redshift range. 

\begin{figure}
	\includegraphics[width=\columnwidth]{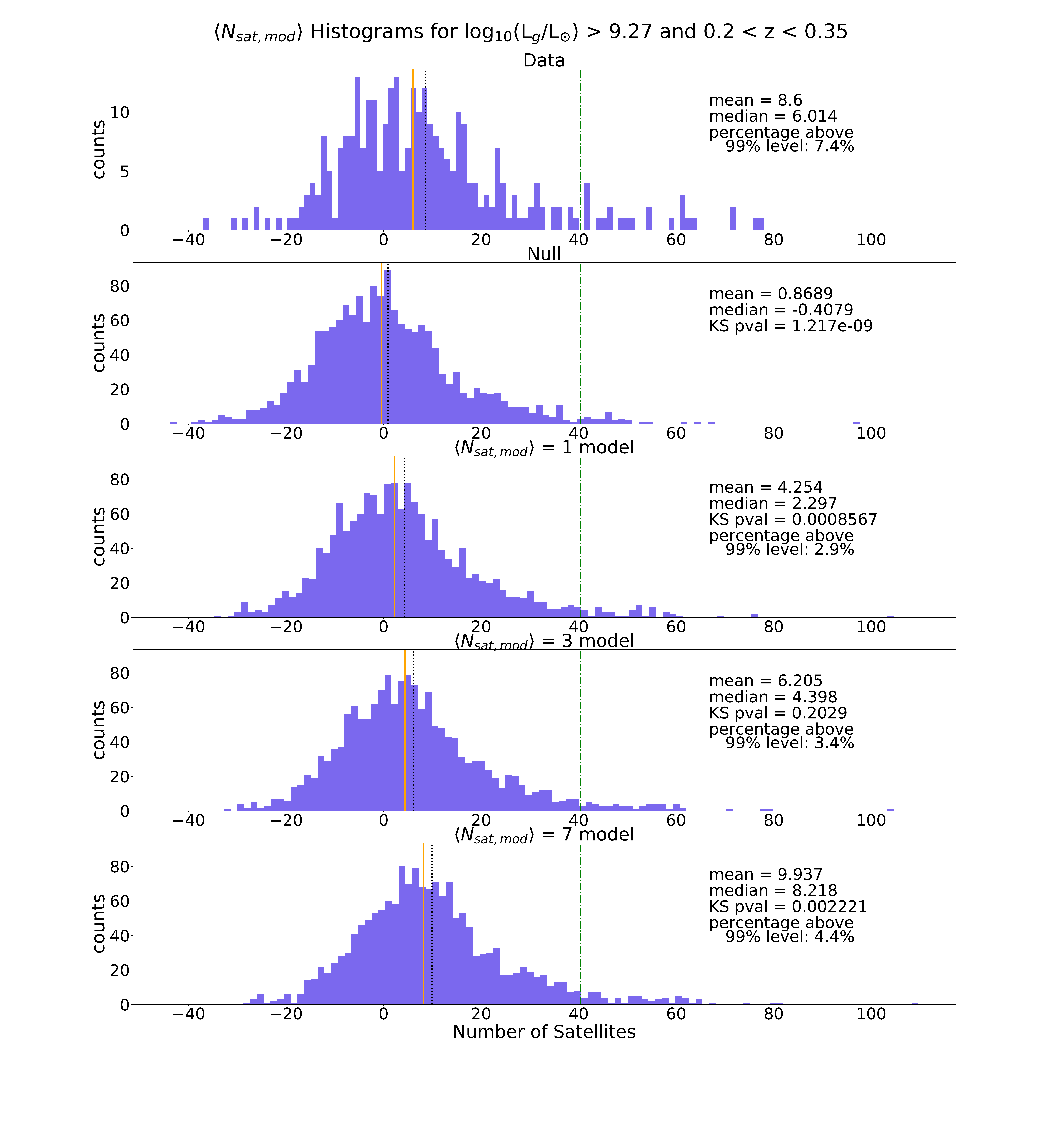}
    \caption{Distributions of the number of satellites found from the data (top panel), the null test (second panel), and models for different intrinsic $\langle N_{sat,mod} \rangle$ numbers for 0.2 $<$ z $<$ 0.35 and log$_{10}$($L$$_{g}$/$L$$_{\odot}$) $\geq$ 9.27 (bottom three panels). The dot-dash line indicates the 99th percent confidence for each bin, as described in Sec. \ref{sec:nsat}. Above this line, the number of satellites would only occur less than 1 percent of the time in a randomly drawn sample of the sky. The null test and the models are described in Sec. \ref{sec:nsat}. $\langle N_{sat,mod} \rangle$ = 1 and 7 illustrate models that are not consistent with the data, with too few and with a median number of satellites that is too low and too high, respectively. $\langle N_{sat,mod} \rangle$ = 3 shows a model that is consistent with the data, as illustrated with a p-value $=$ 0.2. All panels show a skewed distribution, with a tail towards high satellite counts. However, the data and models show a more pronounced tail than the null prediction. The plots also indicate the percentage of LRGs in both the data and the models that exceed the number of satellites at 99 percent confidence.}
    \label{fig:data-null-sats-ex}
\end{figure}

In all redshift and luminosity bins, there is a large scatter in the distribution and a tail toward higher satellite numbers. An example of our N$_{sat}$ distributions can be found in the top panel of Fig. \ref{fig:data-null-sats-ex}. Despite the tail of $N_{sat}$ to high values as seen, e.g. in the top of Fig. \ref{fig:data-null-sats-ex} and Appendix \ref{sec:appdx}, the large spread in $N_{sat}$ makes it difficult to determine at face value how many LRGs have a statistically significant detection of satellites.

To determine the significance of our N$_{sat}$ measurement, we determine how many "satellites" we would expect to detect in a random pointing. This constitutes the null prediction for our measurement.

For each luminosity bin, we randomly select 10,000 galaxies from the EDR. We consider these randomly selected galaxies to be our mock LRGs. We then randomly assign each mock LRG a redshift between $0.2 < z < 0.65$. We then run our statistical background subtraction method on these mock LRGs and apply the selection matrices at the mock redshift. N$_{nn,null}$, N$_{bkg,null}$ and N$_{sat,null}$ are determined in the same way as N$_{nn}$, N$_{bkg}$ and N$_{sat}$ from the real data.

\begin{figure*}
	\includegraphics[scale=0.32]{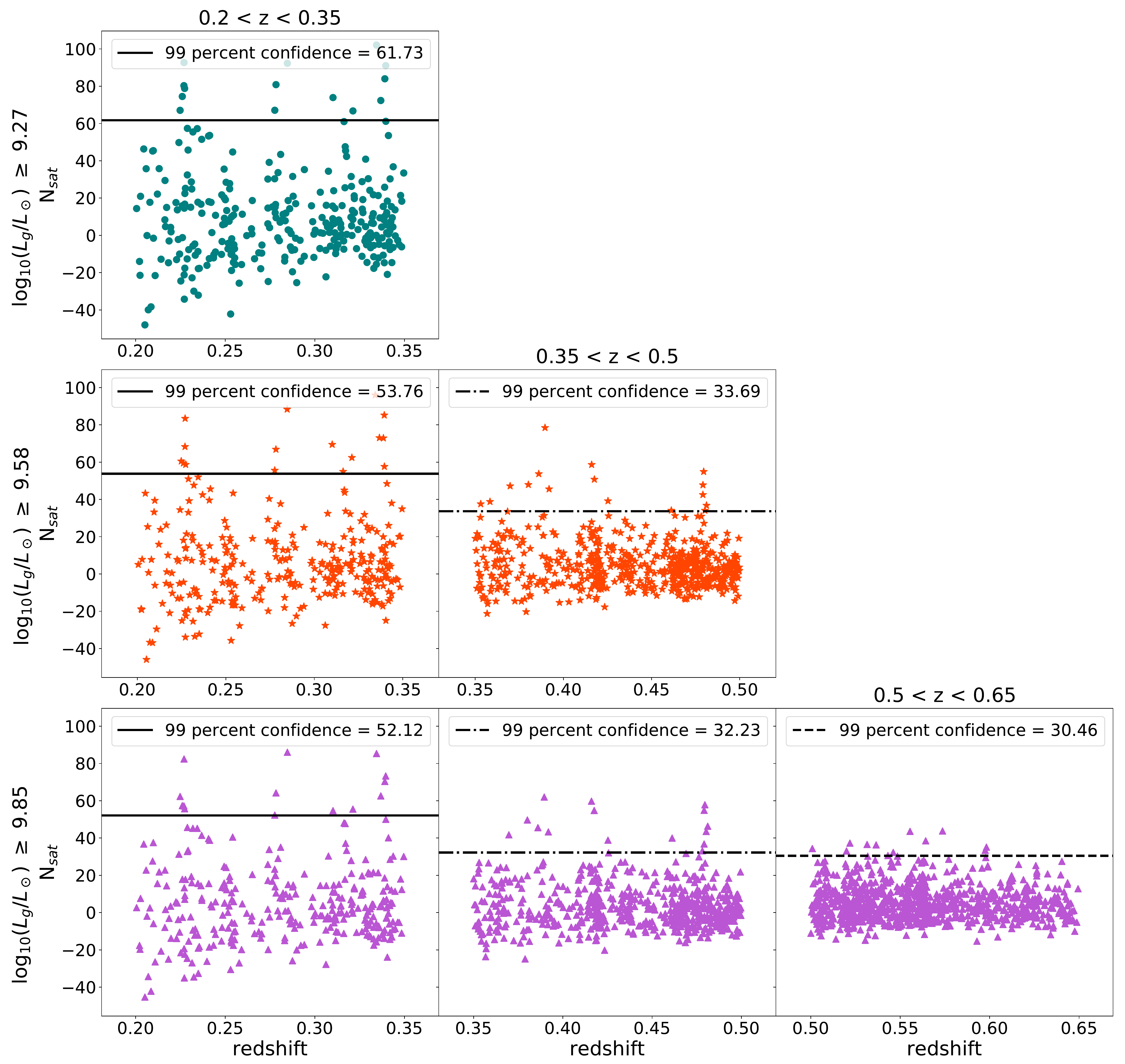}
    \caption{The number of satellites for each LRG plotted against the LRG redshift. The horizontal lines indicate the 99th percent confidence for each bin, as described in Sec. \ref{sec:nsat}. Above this line, the number of satellites would only occur less than 1 percent of the time. Fewer satellites are needed for a significant detection in the high redshift bins compared to the lower redshift bins, at fixed luminosity. Table \ref{tab:summary-table} gives the proportion of LRGs in each bin that have significant satellite detections, as well as the upper and lower bounds given by a 99\% binomial confidence interval. Despite the lower threshold, at fixed luminosity the proportion of LRGs with significant detections decreases with increasing redshift.}
    \label{fig:sats-v-redshift}
\end{figure*}

An example of the distribution of satellites gleaned from the null test is in the second panel from the top of Fig. \ref{fig:data-null-sats-ex} and the results of the null test for all luminosity bins can be found in Figs. \ref{fig:low-model35} through \ref{fig:high-model65}. Results of the null tests for each luminosity-complete sample are used to determine the significance of N$_{sat}$ measurements. We consider there to be a significant satellite detection if the measurement N$_{sat}$ for an individual LRG exceeds the 99th percentile of satellites measured from the null result. The 99th percentile is noted on the plots of satellite numbers throughout this paper.

In Fig. \ref{fig:sats-v-redshift} we show the distribution of $N_{sat}$ vs. redshift for all of our combinations of redshift and luminosity limit. In each panel we denote the 99th percentile for a significant $N_{sat}$ detection as a horizontal line.  The 99th percentile is different for each combination of redshift and luminosity. At fixed luminosity (row) the threshold for a significant detection decreases with increasing redshift. While the 99\% limit decreases towards higher redshift, the fraction of LRGs with a significant number of satellites is constant within the uncertainties, as shown in the third row of Table \ref{tab:summary-table}. This likely stems from the different Poisson uncertainty from the background subtraction in different redshift bins.  On one hand, our constant luminosity threshold will correspond to a brighter $z$-band magnitude limit at lower redshift, which should decrease the number of interlopers.  On the other hand, our color cuts described in Sec. \ref{sec:stat-sub} are very effective at removing high redshift (and likely fainter) interlopers, but reject the lowest fraction of interlopers for our lowest redshift bins and would result in a higher contribution to the uncertainty in $N_{sat}$.  The combination of these effects is likely responsible for the modest dependence of the 99\% confidence level on redshift.

Most LRGs in all bins do not have a significant satellite measurement within a 600 kpc radius. Some LRGs have small negative number of satellites. This is a result of the statistical nature of our subtraction method. The existence of negative satellites - an obviously unphysical phenomenon - is a reflection of the random fluctuations in the background measurement. At times, those fluctuations will produce a high background when compared to near neighbors. 

\begin{figure*}
	\includegraphics[width=\textwidth]{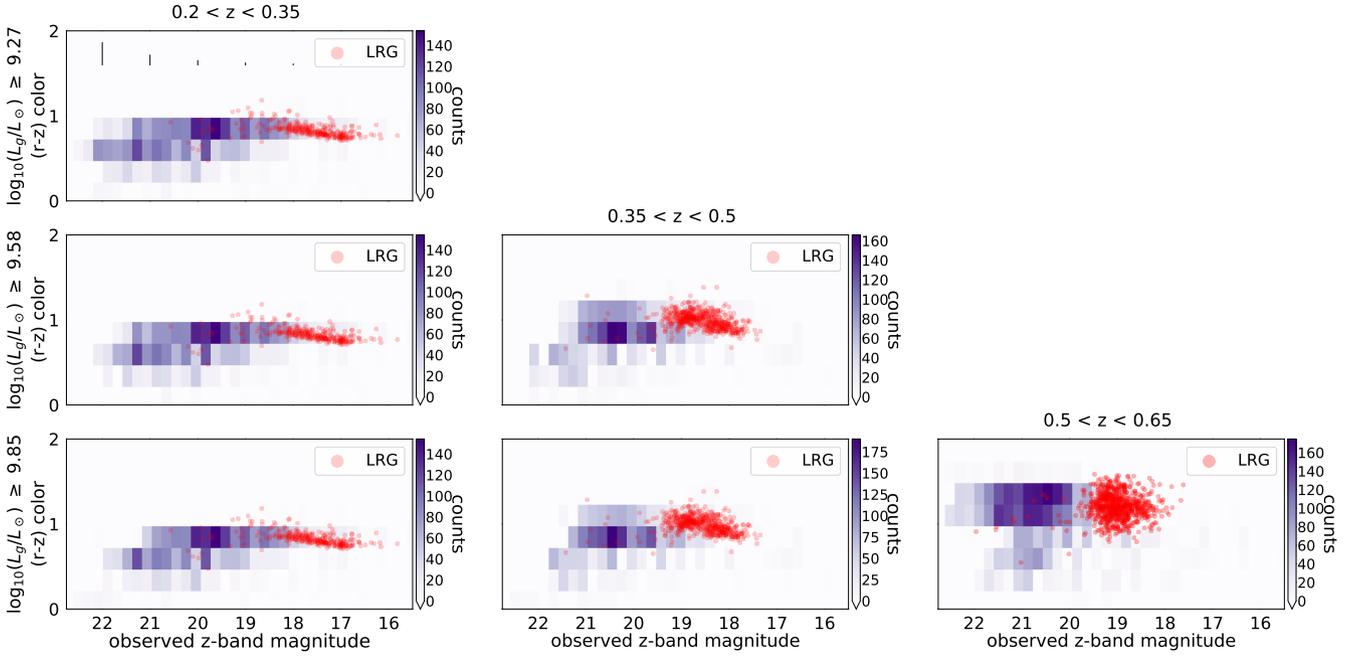}
    \caption{Observed ($r-z$) color vs. $z$-band magnitude diagram for LRGs (red dots) and satellite galaxies (shading) for different redshift and luminosity bins. Each red dot represents an individual LRG in this redshift range and the shaded regions represent the distribution of satellite galaxies in bins of ($r-z$) observed color and observed $z$-band magnitude. The vertical lines across the top of the first panel represents the median uncertainty in the color measurement at the observed $z$-band magnitude at which they are placed. In general, the satellite population is fainter than the population of LRGs, though there is some overlap in the distributions.  In all cases satellite galaxies appear to have ($r-z$) colors similar to or bluer than that of LRGs.  We quantify this in Fig.~\ref{fig:delta-rz}.} 
    \label{fig:ObservedCMD-sats}
\end{figure*}

\begin{figure*}
	\includegraphics[width=\textwidth]{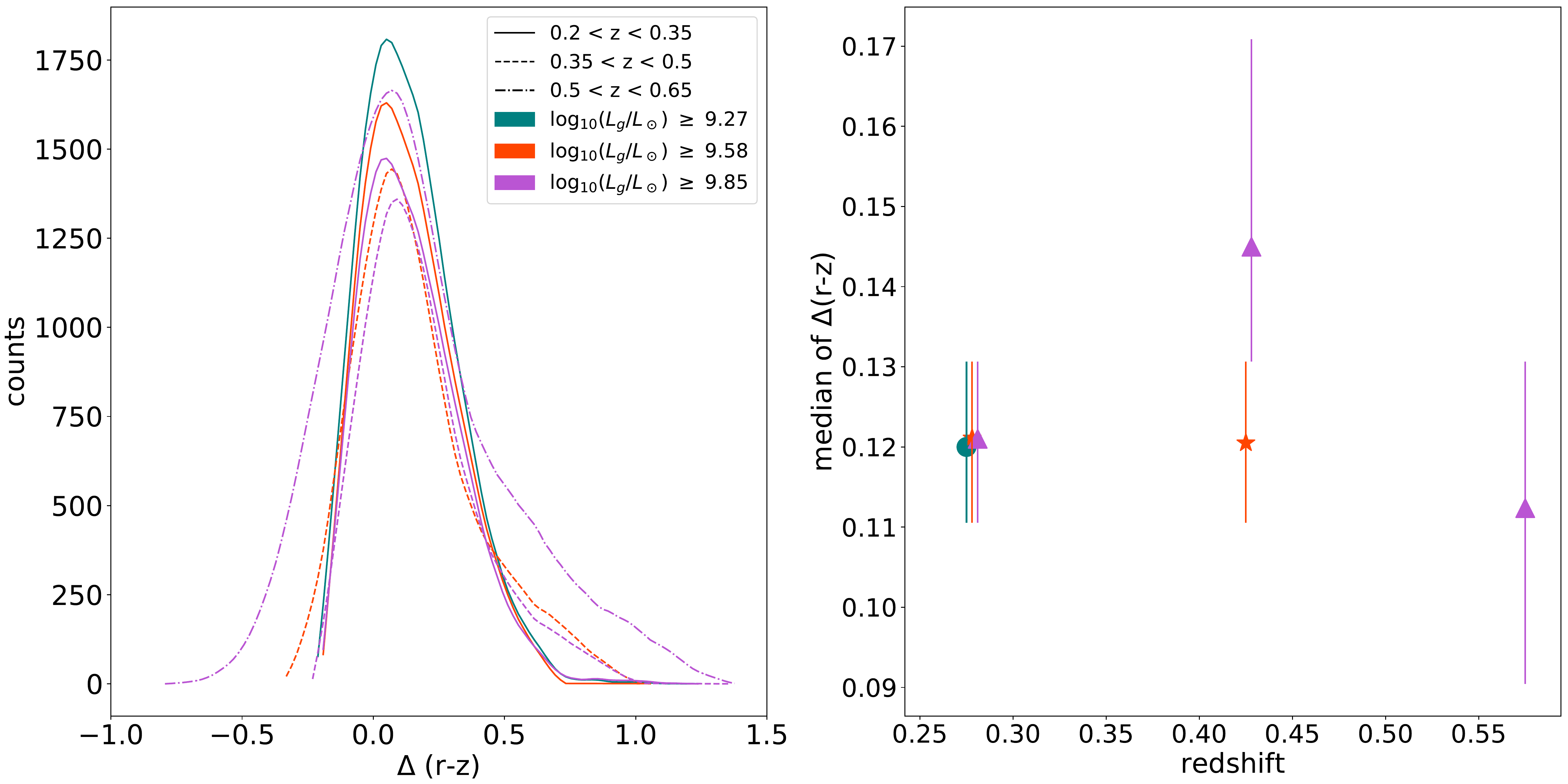}
    \caption{Left: The distribution of the difference between the observed (r-z) colors of satellites and their host LRGs, ($r-z$)$_{LRG}$ - ($r-z$)$_{sat}$ $=$ $\Delta$($r-z$). Colors represent different luminosity limits and line styles represent different redshift ranges. Right: The median value of $\Delta$($r-z$) for each luminosity and redshift bin. Error bars represent 16th to 84th percentile in the weighted median determined through boostrap resampling.}
    \label{fig:delta-rz}
\end{figure*}

\subsection{Color Distribution of LRG Satellites}
\label{sec:color-dist}

While our method makes it impossible to study individual satellite galaxies in detail, it is possible to determine color and magnitude distributions as a population. Fig. \ref{fig:ObservedCMD-sats} illustrates the observed ($r-z$) color vs observed $z$-band magnitude of satellites compared to LRGs in bins of redshift and luminosity. We show the distribution of LRGs and satellites for the full sample. 

In general, the satellite population is fainter than the population of LRGs, though there is some overlap in the distributions. In all cases satellite galaxies appear to have ($r-z$) colors similar to or bluer than that of LRGs.

\begin{figure*}
	\includegraphics[width=\textwidth]{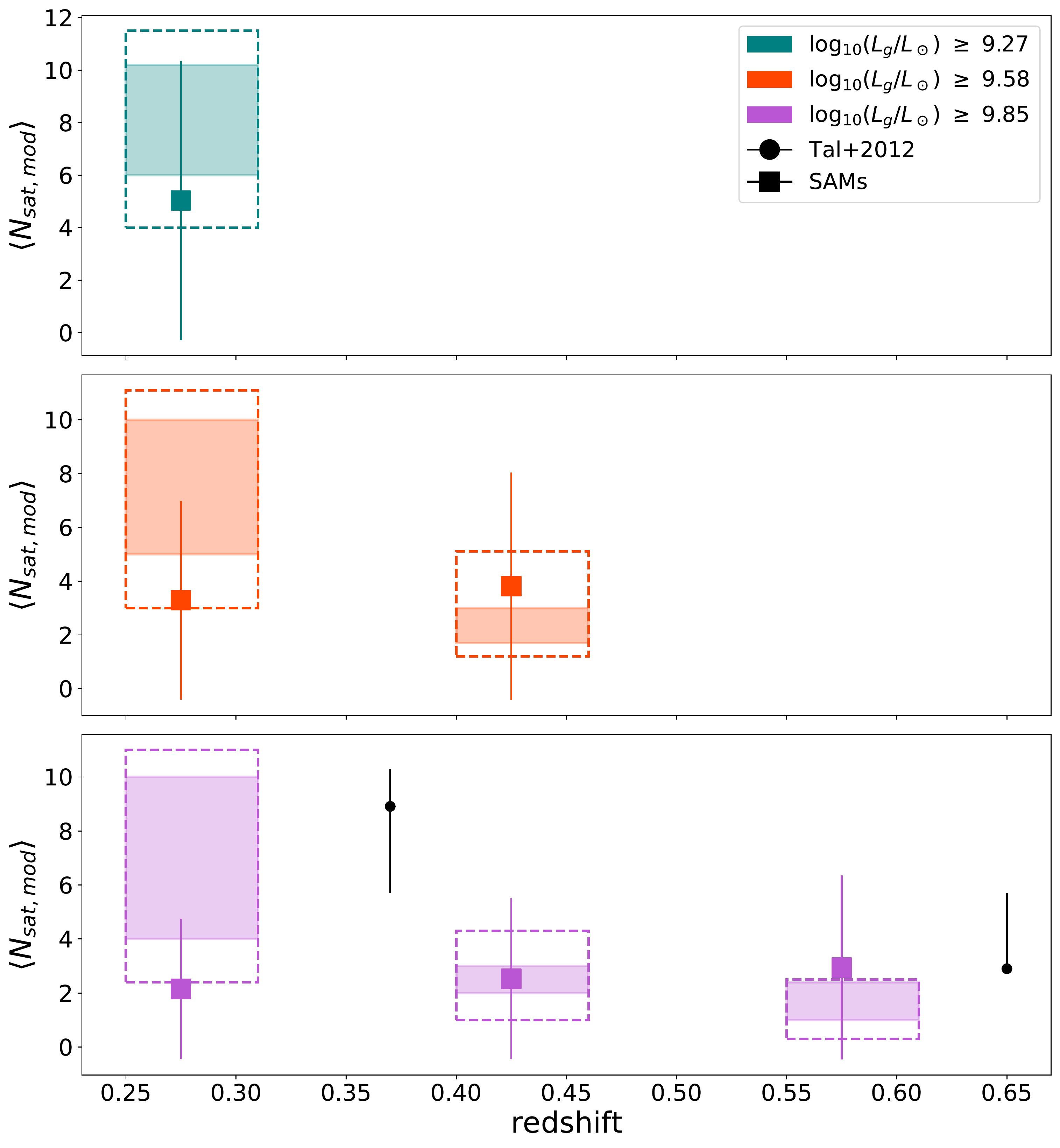}
    \caption{A summary of our forward modelling results, described in Sec. \ref{sec:fwd-model} and Sec. \ref{sec:discuss}. The y-axis is the number of model satellites and the x-axis is redshift. The shaded regions show the range of numbers of satellites with a p $>$ 0.05 and the dashed boxes show the range with a p $>$ 0.003. Each panel includes our forward modeling results as rectangles and results from an analysis of the satellite counts from semi-analytic models of \citet{Hirschmann2016} as square points. The error bars for the SAMs are the 68\% confidence limits on the distribution; the errors in the mean are smaller than the points and are thus not plotted. The bottom panel also includes the satellite numbers for two redshift bins from \citet{Tal2012a}, shown as round black points. These are scaled to our search radius of 600 kpc and the error bars represent the 68\% confidence limits determined by the range of $\alpha$$_{s}$ in \citet{Tal2012a}.}
    \label{fig:model-summary}
\end{figure*}

To quantify the difference in observed ($r-z$) color between LRGs and their satellites, we implement the following process. We cannot identify an individual galaxy as a satellite, hence we do not know the exact satellite ($r-z$) colors. As an approximation, we use the midpoint value of the ($r-z$) color bins as satellite colors. Each ($r-z$) color bin is weighted by the number of galaxies in it. We then subtract this color from the color of the host LRG and we interpolate the resulting $\Delta$($r-z$) onto a reference grid. 

We show the distribution of $\Delta$($r-z$) in the left panel of Fig. \ref{fig:delta-rz} for each redshift and luminosity bin. The right panel shows the medians of the distributions for each luminosity and redshift bin with the error bars representing the 68 percent confidence interval in the weighted median determined by bootstrap resampling. The right panel in Fig. \ref{fig:delta-rz} shows no statistical evidence for redshift evolution.  

\subsection{Forward Modeling the Satellite Population}
\label{sec:fwd-model}

\begin{table*}
	\centering
	\caption{Table summary of satellite count and forward modeling. N$_{LRG}$, N$_{LRG}$(sig. satellites), and N$_{LRG}$(sig. satellites)/N$_{LRG}$ are the total number of LRGs, the total number of LRGs with a significant satellite detection, and the proportion of LRGs with significant satellite counts with binomial confidence intervals, respectively. The row labeled ``$\langle N_{sat,mod} \rangle$ range'' are the range of values of $\langle$N$_{sat,mod}\rangle$ drawn from a Poisson distribution that are consistent with the data to the 95\% confidence (99.75\% confidence).}
	\label{tab:summary-table}
	\begin{tabular}{lccccccc} 
		\hline
		& \multicolumn{7}{c|}{Summary Table of Satellite Counts and Forward Modeling} \\
		\hline
		& \multicolumn{1}{c|}{log$_{10}$(L$_{g}$/L$_{\odot}$) $\geq$ 9.27} 
		 & \multicolumn{2}{c|}{log$_{10}$(L$_{g}$/L$_{\odot}$) $\geq$ 9.58} 
		 &  \multicolumn{3}{c|}{log$_{10}$(L$_{g}$/L$_{\odot}$) $\geq$ 9.85}\\
		& 0.2 $\leq$ z $<$ 0.35 & 0.2 $\leq$ z $<$ 0.35 & 0.35 $\leq$ z $<$ 0.5 &  0.2 $\leq$ z $<$ 0.35 & 0.35 $\leq$ z $<$ 0.5 & 0.5 $\leq$ z $\leq$ 0.65&\\
		\hline
		N$_{LRG}$ &309 & 309 & 617 & 309 & 617 & 897 &\\
		N$_{LRG}$(sig. satellites) & 14 & 16 & 16 & 14 & 15 & 12 &\\
		N$_{LRG}$(sig. satellites)/N$_{LRG}$ & $0.045^{0.085}_{0.020}$ & $0.041^{0.093}_{0.025}$ & $0.026^{0.047}_{0.012}$ &  $0.045^{0.085}_{0.020}$ & $0.024^{0.045}_{0.011}$ & $0.013^{0.027}_{0.006}$ &\\
		$\langle$ N$_{sat,mod}$$\rangle$ range & 6.0-10.2 (4.0-11.5) & 5.0-10.0 (3.0-11.1) & 1.7-3.0 (1.2-4.1) &  4.0-10.0 (2.4-11.0) & 2.0-3.0 (1.0-4.3) & 1.0-2.4 (0.3-2.5) &\\
		\hline
	\end{tabular}
\end{table*}

With the above techniques we can infer the number of LRGs with a statistically significant excess of satellites.  However, because of the large spread in $N_{sat}$ and other potential systematics (see previous sections), we cannot straightforwardly determine the average number of satellites per LRG using only the measured data.  For this reason we develop a method in which we use our null distribution of $N_{sat}$ to forward model the predicted distribution of the measured $N_{sat}$ for an intrinsic distribution of satellites. Modeling in this way attempts to answer the question, if we integrate down the satellite galaxy luminosity function to a certain luminosity and at a certain redshift, would we find that the number distribution of satellite galaxies around LRGs to be consistent with LRGs having an intrinsic number of satellites?

We start with the assumption that the mock LRGs in a given model have an intrinsic mean number of satellites, with the number around each mock LRG independently drawn from a Poisson distribution with the corresponding expectation value.  This number is effectively the integral of the luminosity function of satellites down to the luminosity threshold.  For example, a model with an expectation value of $\langle N_{sat,mod} \rangle$ $=$ 1 would have all mock LRGs populated with a number of satellites drawn from a Poisson distribution with that expectation value.

After we determine how many satellites to add to each mock LRG, we add sources to the near neighbor matrix N$_{nn,null}$ in cells that are at least as bright in the median observed $z$-band magnitude for the color-color cell. In practice, these galaxies are randomly chosen duplicates of galaxies already in those cells, to ensure that the added mock satellites follow the same color-color-magnitude distribution of the actual galaxies. The resulting matrix is N$_{nn,mod}$. The background matrix for the mock LRG remains unmodified. The model number of satellites, N$_{sat,mod}$, is calculated by subtracting N$_{bkg,null}$ from N$_{nn,mod}$, and then applying the appropriate selection matrix.

A sample of our model results is shown in Fig. \ref{fig:data-null-sats-ex}, which indicates the mean, median, and p-value from the Komogorov-Smirnov 2-sample test. The p-value indicates whether or not the model and data are consistent with being drawn from the same distribution. Small p-values indicate that the model and the data likely did not come from the same distribution. For example, the case shown in Fig. \ref{fig:data-null-sats-ex} for 0.2 $<$ z $<$ 0.35 and log$_{10}$($L$$_{g}$/$L$$_{\odot}$) $\geq$ 9.27 has an acceptable p-value ($>0.003$) for the model with $\langle$N$_{sat,mod}\rangle$ = 3, but significantly lower p-values for all other $\langle$N$_{sat,mod}\rangle$, including the null result of zero mean satellites. The remaining comparisons between the data, null result, and models can be found in Appendix \ref{sec:appdx}.

Fig. \ref{fig:model-summary} shows a summary of our model results. The solid boxes represent the range of satellite numbers expected for p $>$ 0.05 and the dashed boxes represent the range for p $>$ 0.003. The colors represent the same luminosity bins as Fig. \ref{fig:sats-v-redshift}. Based on these results, we can rule out a lack of evolution to $z<0.5$ at a fixed luminosity limit at the greater than 95\% level.  We conclude this as the allowed range of  $\langle$N$_{sat,mod}\rangle$ at $z<0.5$ in each luminosity bin do not overlap.  We cannot reject the no-evolution hypothesis at the greater than 99.7\% level as the allowed values at that confidence level overlap across all redshifts.

We do not have the precision to constrain the faint end of the luminosity function, but the lack of a clear increase in $\langle$N$_{sat,mod}\rangle$ with decreasing luminosity limit in Fig. \ref{fig:model-summary} implies a flat or declining faint-end luminosity slope for satellites. A summary of these result can be found in Table \ref{tab:summary-table}.


\section{Discussion}
\label{sec:discuss}

\subsection{Colors and Numbers of LRG satellites}
\label{satcolors}

While our statistical subtraction methods leaves us unable to study any single satellite galaxy in particular, nevertheless we are able to study satellite properties on a population level.

Fig. \ref{fig:delta-rz} illustrates the difference in the color distribution between satellites and their host LRG. The color difference across all bins are consistent with each other. The color distribution for the highest redshift and highest luminosity bin has a significantly larger spread than the other bins and extends to fainter luminosities.  The origin of this is seen in the bottom right-hand panel of Fig. \ref{fig:ObservedCMD-sats} in which we see that the LRGs and the reddest satellites are significantly redder than in lower redshift bins, though the satellites still reach to as blue colors as in the other redshift bins.  This behavior stems from the color-dependent k-corrections, i.e. the observed $(r-z)$ color of passive galaxies becomes rapidly redder with increasing redshift while the observed color for blue galaxies changes much less.  The result is an increase in the difference between the reddest and bluest objects in observed color as one looks to higher redshifts.

As shown in Fig.~\ref{fig:model-summary}, at a fixed redshift we see substantial overlap between acceptable models for each luminosity limit, such that we cannot rule that the mean number of satellites is independent of the luminosity limit, at least over our luminosity range.  Despite that, there are some hints of a dependence on the luminosity limit.  In the low redshift bin we can see that the $2-$ and $3-\sigma$ lower bounds of acceptable models increases as the luminosity limit decreases. For log$_{10}$($L$$_{g}$/$L$$_{\odot}$) $\geq$ 9.27, the lower bound is $\langle N_{sat,mod}\rangle = 4$. The lower $3-\sigma$ bound for log$_{10}$($L$$_{g}$/$L$$_{\odot}$) $\geq$ 9.58 is $\langle N_{sat,mod}\rangle = 3$ and for log$_{10}$($L$$_{g}$/$L$$_{\odot}$) $\geq$ 9.85 $\langle N_{sat,mod}\rangle = 2.4$.  This is somewhat as expected, as we should be increasing the number of satellites as we integrate further down the luminosity function.

As described in Sec. \ref{sec:fwd-model}, there is tentative evidence for evolution at $z<0.5$ in satellites selected to a fixed luminosity.  Interpreting this evolution in terms of the true number of satellites down to a given mass limit is complicat ed, as satellites will evolve in $M/L_\star$ over time and as such should on average be reducing their luminosity.  This implies that satellites which are just visible above our luminosity limit at higher redshift may fade below our limit by a lower redshift.  The tentative evolution we see may therefore be indicative of an even stronger evolution in the number of satellites to a fixed stellar mass threshold.  We will address this issue in a future paper.

We compare our results to \citet[][hereafter T12]{Tal2012a} by integrating the luminosity function they found for satellites. The T12 analysis differs from ours in important ways. First, they only go down to a luminosity of log$_{10}$($L$$_{g}$/$L$$_{\odot}$) $\sim$ 10 at 0.28 $<$ z $<$ 0.4 and log$_{10}$($L$$_{g}$/$L$$_{\odot}$) $\sim$ 10.5 at 0.6 $<$ z $<$ 0.7, whereas this work goes down to log$_{10}$($L$$_{g}$/$L$$_{\odot}$ $\sim$ 9.85 out to $z=0.65$. Second, T12 measures satellites out to 1 Mpc, which is a larger radius than we search in this work.  To account for these differences we first integrate the luminosity function from T12 down to their luminosity limits.  These limits are brighter than ours but from a small sample at lower redshift T12 showed that the luminosity function starts flattening below log$_{10}$($L$$_{g}$/$L$$_{\odot}\sim 10$.  We therefore do not think that the T12 satellite numbers when integrated to our luminosity limits would significantly increase.   We then account for the different in radial search areas by scaling the T12 result by the ratio of the mean N$_{sat}$ at 0.6 Mpc and 1 Mpc from our own data

The T12 points are plotted in Fig. \ref{fig:model-summary}. The error bars represent the 68\% confidence limits determined by the range of $\alpha$$_{s}$ in T12.  The T12 points are in broad agreement with our data though they appear to have a slightly higher number of satellites.

Additionally, we also compare our results to semi-analytic models based on \cite[][private communication]{Hirschmann2016}\footnote{https://sites.google.com/inaf.it/gaea/?pli=1}. We measure the number of satellites around  simulated LRGs out to a 3D radius of 600 kpc in bins of redshift and $g$-band luminosity. We plot these model predictions in Fig. \ref{fig:model-summary}. There is good agreement between our results and the semi-analytic models.  At a fixed redshift, the SAM has the same trend with luminosity limit as in our data.  However, they show a shallower trend with redshift at a fixed luminosity limit than does our data.  It is not clear if this difference lies in the approximations we make in our comparison, e.g. the use of 3D radii for the models, or if it reflects a true problem with the simulations.  We will treat this in more depth in a future work. 

\subsection{The Future Mass Growth of LRGs}
\label{massgrowth}

We next estimate the amount of mass that can be gained by each LRG resulting from satellite accretion at $z<0.65$.  We cannot measure the mass of individual satellite galaxies, but can use their statistical color and luminosity distribution relative to that of the host LRGs to compute the average amount of mass in satellites. We start with the $\Delta(r-z)$ and $\Delta$zmag, the average difference between the observed $z$-band magnitude for LRGs and satellites. We use EZgal \citep[][]{Mancone2012} to output a model from \citet{Conroy09} with a Chabrier initial mass function \citep[][]{Chabrier2003}.  We used an exponentially declining SFH from this model to define a relationship between observed $(r-z)$ color and the ratio of the stellar mass ($M_\star$) to the observed $z$-band flux ($f_z$).  This was done at the mean observed redshift $z_{obs}$ of each of our LRG subsamples.  In this model the relation between color and log$_{10}(M_\star/f_z)$ represents a trend with age.  This relation is valid for all smooth SFHs and dust extinction moves galaxies relatively parallel to this relation.  Using the mean difference in observed color between LRGs and their satellites $\Delta(r-z)$, we use this relation to infer a difference in log$_{10}(M_\star/f_z)$ between LRGs and satellites, which we call $\Delta$ log$_{10}(M_\star/f_z)$.  From this difference and the ratio of the mean $z$-band flux between LRGs and satellites we can compute the ratio of the stellar mass in the two populations as

\begin{equation}
   log_{10} \left( \frac{M_{\star,LRG}}{M_{\star,sat}} \right) = \Delta log_{10} \left( \frac{M_{\star}}{f_z} \right) - log_{10} \left( \frac{f_{z,sat}}{f_z,LRG} \right).
\end{equation}

We make a number of assumptions and simplifications in this calculation. First, we assume all satellites have the same star formation history or can be represented by a single star formation history. In reality, the satellites with lower luminosity may have different star formation histories but at the same time all smooth SFHs should lie in the same plane of log$_{10}(M_{\star}/f_z)$ vs. $(r-z)_{obs}$. Additionally, we assume that the star formation histories of LRGs and satellites are the same and we assume that the effects of dust are negligible on the relation we use.

We find that log$_{10} \left( \frac{M_{\star,LRG}}{M_{\star,sat}} \right)$ ranges from 0.77 to 1.22, with the highest mass in satellites occurring in our highest redshift and highest luminosity limit bin.  The lowest mass in satellites occurs in our lowest redshift and lowest luminosity limit bin.  At a given redshift, log$_{10} \left( \frac{M_{\star,LRG}}{M_{\star,sat}} \right)$ is constant to within $\lesssim$ 0.1 dex for satellites with different luminosity limits. Furthermore, at the same luminosity limit, but over different redshifts, log$_{10} \left( \frac{M_{\star,LRG}}{M_{\star,sat}} \right)$ increases as redshift increases. The increase in relative satellite mass is $\sim 0.2$ dex between the two lowest redshift bins for the log$_{10}(L_g)>9.58$ limit and $\sim 0.3$ dex between the two redshift lowest bins for the log$_{10}(L_g)>9.85$ limit. There is no increase in relative satellite mass between the middle and highest redshift bin for the log$_{10}(L_g)>9.85$ limit.

There are two reasons why the relative mass in satellites might increase to higher redshift. First, it could be that there are more satellites at higher redshift and that the rate of consumption by the LRGs overtakes the accumulation of new satellites. Second, it could be that all satellites will move to lower mass-to-light ratio values toward higher redshift because they are younger, so a fixed luminosity limit will probe to a lower satellite stellar mass limit at higher redshift. We cannot distinguish between these two possibilities without measuring the rest-frame colors of satellites or their stellar mass. However, given these caveats we can say that LRGs can at most grow by $\sim 15\%$ from $z=0.575$ to the present day and by $\sim 6\%$ from $z=0.275$ to the present day. This is consistent with the findings from \cite{Brown2008}, which determined that the mass of massive red galaxies grows by 30\% since z $<$ 1.

\section{Summary and Conclusions}
\label{sec:concl}

In this study, we use the deep photometry from the DESI Legacy Imaging Surveys to characterize the satellite population around SDSS-identified luminous red galaxies. SDSS LRGs represent a homogeneous population of massive red ellipticals, ideal for investigating the mechanisms by which the mass of massive ellipticals builds up. We use the Legacy Survey photometry for SDSS DR14 LRGs froom the LOWZ and CMASS samples. The deeper Legacy Survey photometry allows us to carry out a characterization of LRG satellites to fainter luminosities and higher redshifts than previously possible for a large sample. We have 1,823 LRGs in our 25 square degree sample area.

We perform a statistical background subtraction in ($g-r$), ($r-z$), and $z$-band magnitude space. To determine what counts as a significant satellite detection, an LRG must have a number of satellites above the 99th percentile of the null result in each luminosity bin. We develop a technique to forward model the satellite distribution, accounting for  the survey selection function and systematics in our background subtraction technique.  This method deals with the lack of a priori knowledge of the redshifts and luminosities of the satellite galaxies.

The main results of this paper are as follows:

\begin{itemize}
  
 \item LRGs are in general brighter than their satellites in the observed $z$-band, but satellite galaxies have similar or slightly bluer observed ($r-z$) colors to LRGs. This is illustrated in Figs. \ref{fig:ObservedCMD-sats} and \ref{fig:delta-rz}.
 
 \item We show in Fig. \ref{fig:sats-v-redshift} and in Table \ref{tab:summary-table} that the proportion of LRGs that have significant satellite detections decreases with increasing redshift. This does not have a straightforward interpretation with regards to the true population of satellites, as the threshold for significance varies with redshift and luminosity.  
  \item To better interpret the number of satellites around the mean LRG in each bin we developed a forward modeling approach in which we use mock samples to simulate the observed signature of difference satellite populations, assuming that the total number of satellites is drawn from a Poisson distribution with expectation value $\langle {\rm~N}_{sat,mod}\rangle$. For LRGs at all combinations of redshift and rest-frame $g$-band satellite luminosity limit, we can strongly rule out a lack of satellites (Fig. \ref{fig:model-summary}).  
  

 \item At fixed redshift, the 3-$\sigma$ lower bounds go toward lower acceptable $N_{sat,mod}$. There is, however, significant overlap in acceptable models. This implies a flat or declining faint-end luminosity slope for satellites.

\item In our two highest luminosity bins we find tentative evidence for evolution in $\langle {\rm~N}_{sat,mod}\rangle$ out to $z<0.5$ though cannot rule out the no-evolution hypothesis as the greater than 99.7\% level.  The interpretation of this evolution is complicated by the expected $M/L_\star$ evolution of satellite galaxies over this redshift range.
 
 
 \item We calculated that LRGs can at most grow by $\sim 15\%$ from $z=0.575$ to the present day and by $\sim 6\%$ from $z=0.275$ to the present day. 

\end{itemize}

This pilot program has demonstrated the ability of the DESI Legacy Imaging Survey, coupled with LRG spectroscopy, to probe the satellite properties of the most massive galaxies in the Universe.  Despite the larger area of the EDR, it is not practical in our sample to investigate the trends of satellite galaxies as a function of LRG properties such as stellar mass.  In a future work we will apply these techniques to the full DESI Legacy Imaging Survey, in order to analyze the full population of SDSS-identified LRGs and to investigate the luminosity function of satellite galaxies and their dependence on LRG property.

\section*{Acknowledgements}

The authors would like to thank John Moustakas at Siena College for calculating the masses of LRGs used in this project. We thank Arjun Dey and Stephanie Juneau for useful discussions that helped to spawn this project. The authors acknowledge the support of  NSF AST-1716690, AST-1517815, NASA ADAP-80NSSC19K0592, ADAP-NNX17AF25G, NASA HST-AR-14310.001-A.  They also acknowledge the support of the International Space Sciences Institute in Bern, which hosted discussions of this work as part of the “The Effect of Dense Environments on Gas in Galaxies over 10 Billion Years of Cosmic Time” and “COSWEB: The Cosmic Web and Galaxy Evolution” teams. M.T. acknowledges the support of a University Graduate Fellowship and the Lowry Graduate Fellowship from the University of Kansas. G.R. thanks the Lorentz Center in Leiden for support via their workshop “Galaxy Evolution in the Cosmic Web.  G.R, acknowledges the support of an ESO visiting science fellowship.

The Legacy Surveys consist of three individual and complementary projects: the Dark Energy Camera Legacy Survey (DECaLS; Proposal ID \#2014B-0404; PIs: David Schlegel and Arjun Dey), the Beijing-Arizona Sky Survey (BASS; NOAO Prop. ID \#2015A-0801; PIs: Zhou Xu and Xiaohui Fan), and the Mayall z-band Legacy Survey (MzLS; Prop. ID \#2016A-0453; PI: Arjun Dey). DECaLS, BASS and MzLS together include data obtained, respectively, at the Blanco telescope, Cerro Tololo Inter-American Observatory, NSF’s NOIRLab; the Bok telescope, Steward Observatory, University of Arizona; and the Mayall telescope, Kitt Peak National Observatory, NOIRLab. The Legacy Surveys project is honored to be permitted to conduct astronomical research on Iolkam Du’ag (Kitt Peak), a mountain with particular significance to the Tohono O’odham Nation.

NOIRLab is operated by the Association of Universities for Research in Astronomy (AURA) under a cooperative agreement with the National Science Foundation.

This project used data obtained with the Dark Energy Camera (DECam), which was constructed by the Dark Energy Survey (DES) collaboration. Funding for the DES Projects has been provided by the U.S. Department of Energy, the U.S. National Science Foundation, the Ministry of Science and Education of Spain, the Science and Technology Facilities Council of the United Kingdom, the Higher Education Funding Council for England, the National Center for Supercomputing Applications at the University of Illinois at Urbana-Champaign, the Kavli Institute of Cosmological Physics at the University of Chicago, Center for Cosmology and Astro-Particle Physics at the Ohio State University, the Mitchell Institute for Fundamental Physics and Astronomy at Texas A\&M University, Financiadora de Estudos e Projetos, Fundacao Carlos Chagas Filho de Amparo, Financiadora de Estudos e Projetos, Fundacao Carlos Chagas Filho de Amparo a Pesquisa do Estado do Rio de Janeiro, Conselho Nacional de Desenvolvimento Cientifico e Tecnologico and the Ministerio da Ciencia, Tecnologia e Inovacao, the Deutsche Forschungsgemeinschaft and the Collaborating Institutions in the Dark Energy Survey. The Collaborating Institutions are Argonne National Laboratory, the University of California at Santa Cruz, the University of Cambridge, Centro de Investigaciones Energeticas, Medioambientales y Tecnologicas-Madrid, the University of Chicago, University College London, the DES-Brazil Consortium, the University of Edinburgh, the Eidgenossische Technische Hochschule (ETH) Zurich, Fermi National Accelerator Laboratory, the University of Illinois at Urbana-Champaign, the Institut de Ciencies de l’Espai (IEEC/CSIC), the Institut de Fisica d’Altes Energies, Lawrence Berkeley National Laboratory, the Ludwig Maximilians Universitat Munchen and the associated Excellence Cluster Universe, the University of Michigan, NSF’s NOIRLab, the University of Nottingham, the Ohio State University, the University of Pennsylvania, the University of Portsmouth, SLAC National Accelerator Laboratory, Stanford University, the University of Sussex, and Texas A\&M University.

BASS is a key project of the Telescope Access Program (TAP), which has been funded by the National Astronomical Observatories of China, the Chinese Academy of Sciences (the Strategic Priority Research Program “The Emergence of Cosmological Structures” Grant \#XDB09000000), and the Special Fund for Astronomy from the Ministry of Finance. The BASS is also supported by the External Cooperation Program of Chinese Academy of Sciences (Grant \#114A11KYSB20160057), and Chinese National Natural Science Foundation (Grant \#11433005).

The Legacy Survey team makes use of data products from the Near-Earth Object Wide-field Infrared Survey Explorer (NEOWISE), which is a project of the Jet Propulsion Laboratory/California Institute of Technology. NEOWISE is funded by the National Aeronautics and Space Administration.

The Legacy Surveys imaging of the DESI footprint is supported by the Director, Office of Science, Office of High Energy Physics of the U.S. Department of Energy under Contract No. DE-AC02-05CH1123, by the National Energy Research Scientific Computing Center, a DOE Office of Science User Facility under the same contract; and by the U.S. National Science Foundation, Division of Astronomical Sciences under Contract No. AST-0950945 to NOAO.

Some of the results in this paper have been derived using the healpy and HEALPix packages.

\section*{Data Availability}

The data analyzed in this paper come from DR8 of the DESI Legacy Imaging Survey. These data are available to the public at https://www.legacysurvey.org/dr8/. Spectroscopic data is from DR14 of the Slone Digitial Sky Survey and can be accessed at https://www.sdss.org/dr14/spectro/spectro\_access/.



\bibliographystyle{mnras}
\bibliography{LRGproject} 




\appendix

\section{Satellite Distribution Modeling}
\label{sec:appdx}

\begin{figure*}
	\includegraphics[width=0.8\textwidth]{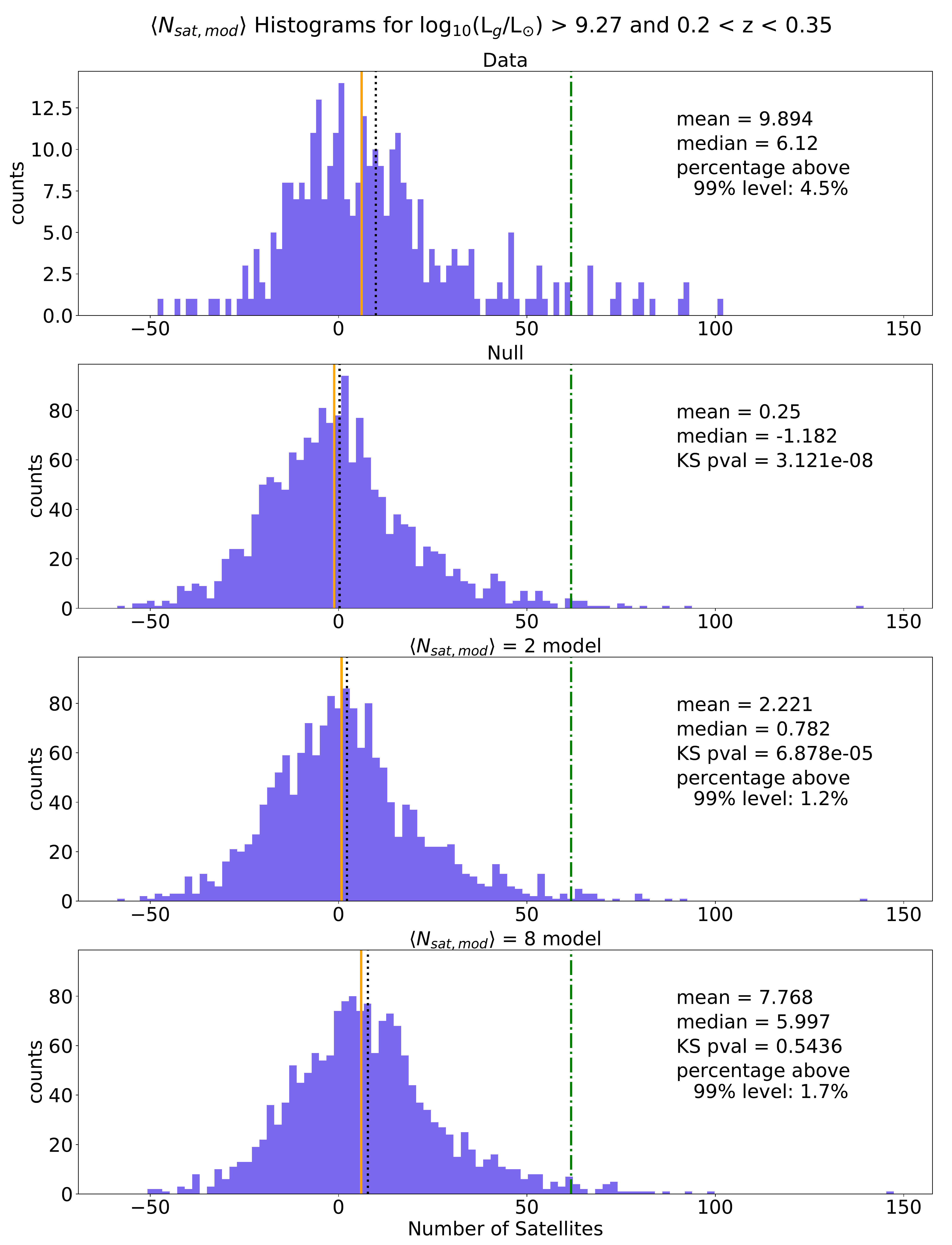}
    \caption{Distributions of the number of satellites found from the data, the null test, and examples of Poisson models for 0.2 $<$ z $<$ 0.35 and log$_{10}$(L) $>$ 9.27. A description of how models are formulated can be found in the text. The dot-dash line indicates 99 percent confidence. The distribution the data is compared to the null test and the models. The model $\langle N_{sat,mod} \rangle = 8$ is included as an example of a model that is a good fit to the data, according to the 2-sample K-S test, and $\langle N_{sat,mod} \rangle = 2$ is included as an example of model that does not have a good fit to the data.}
    \label{fig:low-model35}
\end{figure*}

\begin{figure*}
	\includegraphics[width=0.8\textwidth]{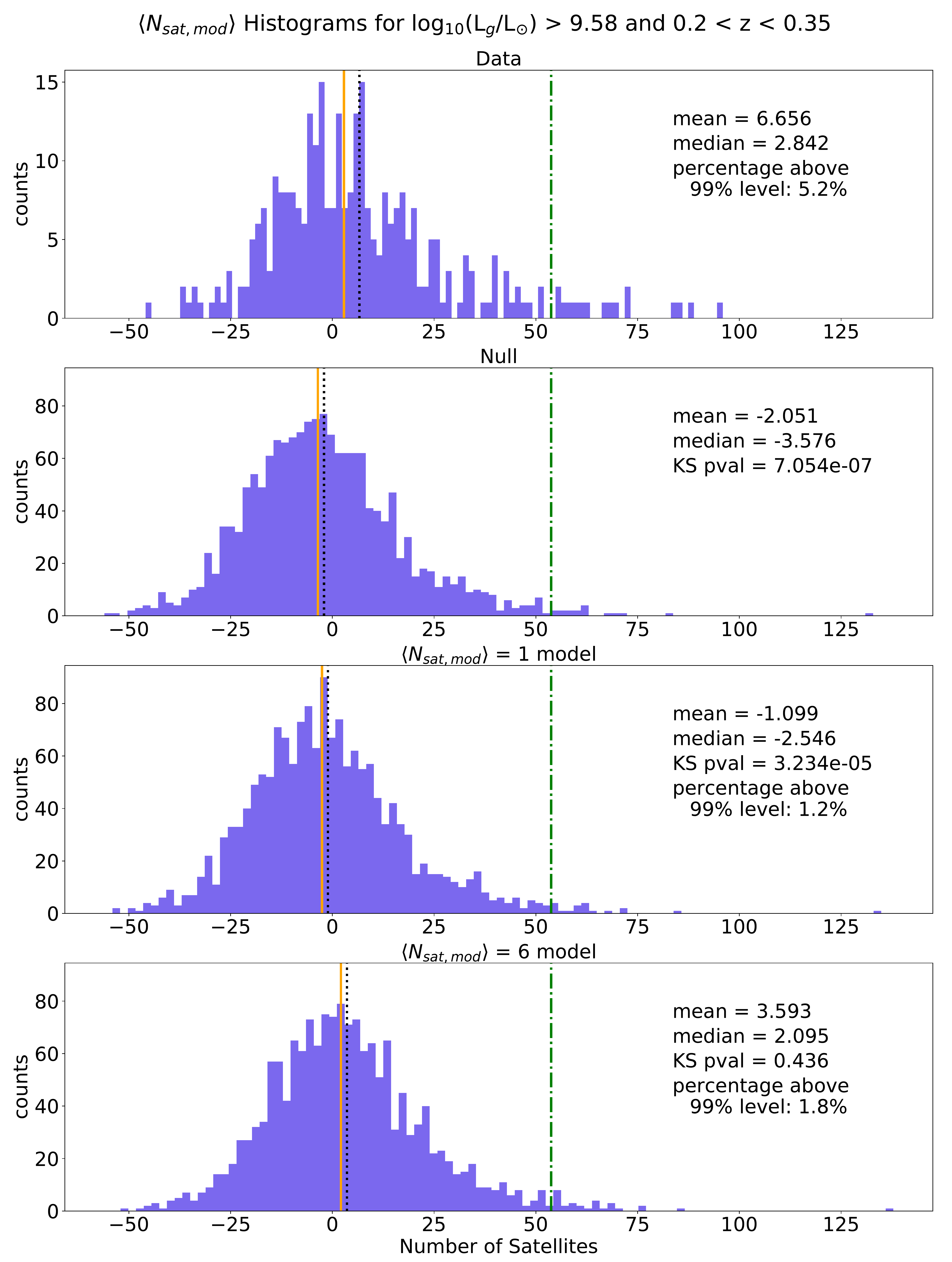}
    \caption{Distributions of the number of satellites found from the data, the null test, and Poisson models for 0.2 $<$ z $<$ 0.35 and log$_{10}$(L) $>$ 9.58. A description of how models are formulated can be found in the text. The dot-dash line indicates the 99 percent confidence. The distribution the data is compared to the null test and the models. The model $\langle N_{sat,mod} \rangle = 6$ is included as an example of a model that is a good fit to the data, according to the 2-sample K-S test, and $\langle N_{sat,mod} \rangle = 1$ is included as an example of model that does not have a good fit to the data.}
    \label{fig:mid-model35}
\end{figure*}

\begin{figure*}
	\includegraphics[width=0.8\textwidth]{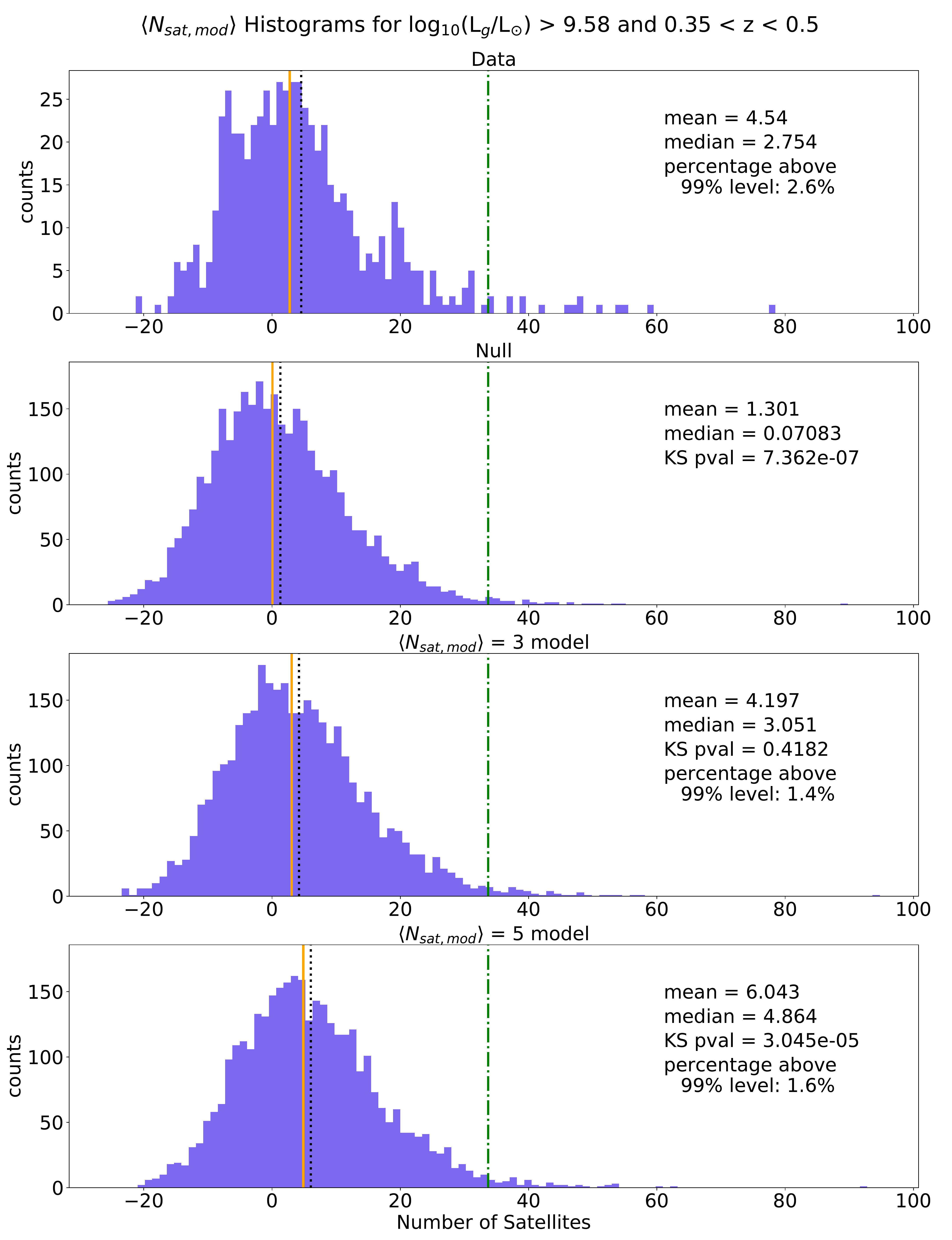}
    \caption{Distributions of the number of satellites found from the data, the null test, and Poisson models for 0.35 $<$ z $<$ 0.5 and log$_{10}$(L) $>$ 9.58. A description of how models are formulated can be found in the text. The dot-dash line indicates the 99 percent confidence. The distribution the data is compared to the null test and the models. The model $\langle N_{sat,mod} \rangle = 3$ is included as an example of a model that is a good fit to the data, according to the 2-sample K-S test, and $\langle N_{sat,mod} \rangle = 5$ is included as an example of model that does not have a good fit to the data.}
    \label{fig:mid-model05}
\end{figure*}

\begin{figure*}
	\includegraphics[width=0.8\textwidth]{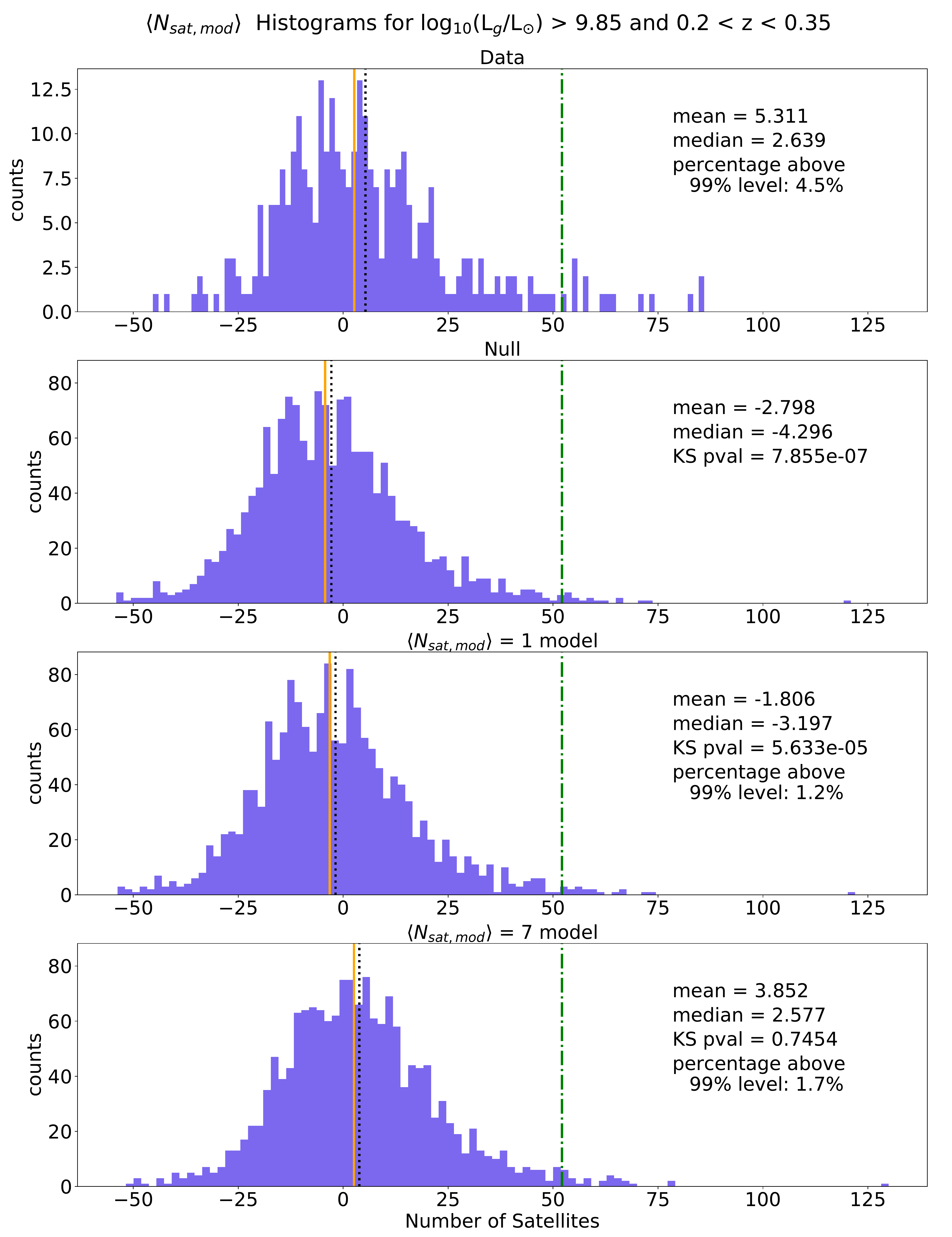}
    \caption{Distributions of the number of satellites found from the data, the null test, and Poisson models for 0.2 $<$ z $<$ 0.35 and log$_{10}$(L) $>$ 9.85. A description of how models are formulated can be found in the text. The dot-dash line indicates the 99 percent confidence. The distribution the data is compared to the null test and the models. The model $\langle N_{sat,mod} \rangle = 7$ is included as an example of a model that is a good fit to the data, according to the 2-sample K-S test, and $\langle N_{sat,mod} \rangle = 1$ is included as an example of model that does not have a good fit to the data.}
    \label{fig:high-model35}
\end{figure*}

\begin{figure*}
	\includegraphics[width=0.8\textwidth]{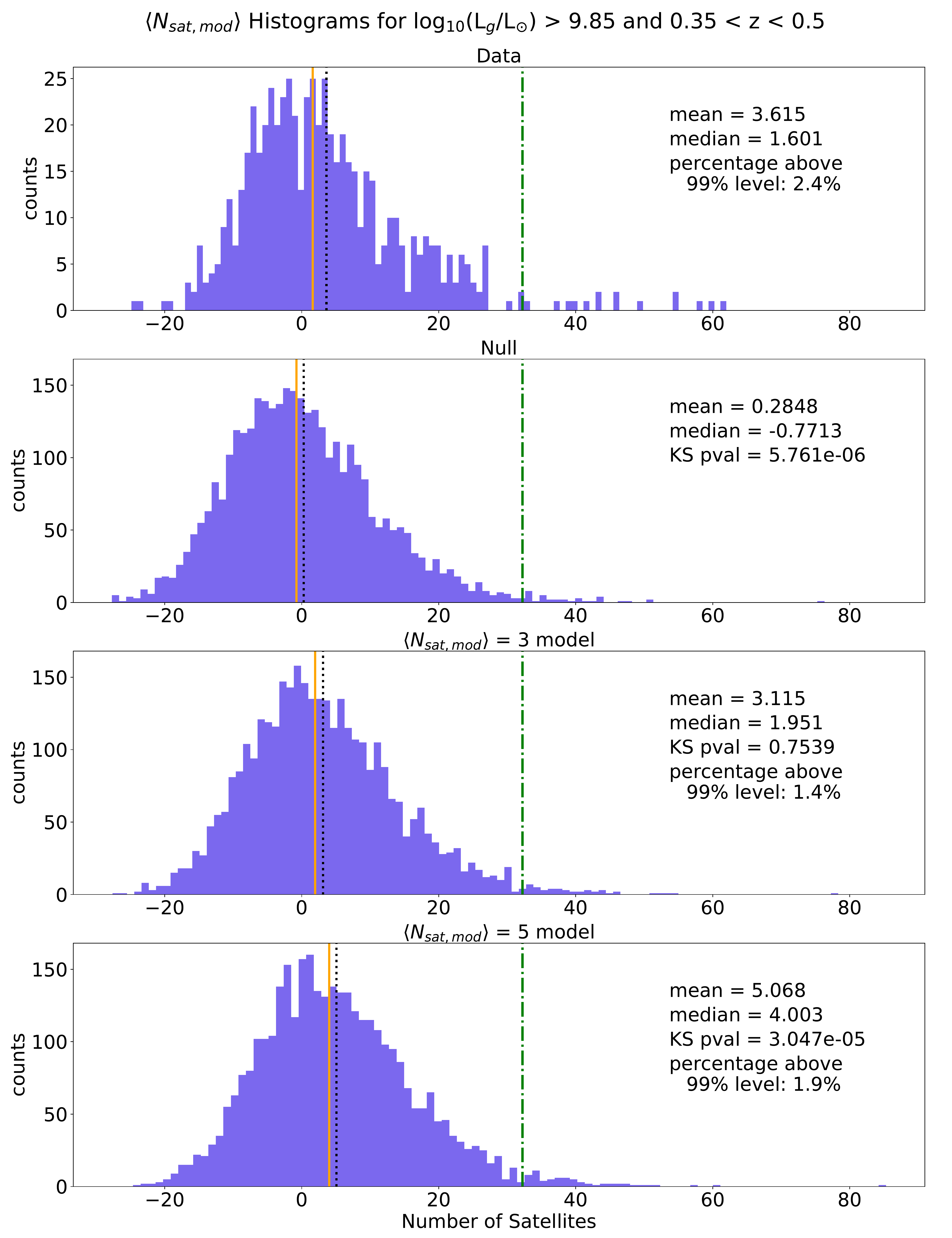}
    \caption{Distributions of the number of satellites found from the data, the null test, and Poisson models for 0.35 $<$ z $<$ 0.5 and log$_{10}$(L) $>$ 9.85. A description of how models are formulated can be found in the text. The dot-dash line indicates the 99 percent confidence. The distribution the data is compared to the null test and the models. The model $\langle N_{sat,mod} \rangle = 3$ is included as an example of a model that is a good fit to the data, according to the 2-sample K-S test, and $\langle N_{sat,mod} \rangle = 5$ is included as an example of model that does not have a good fit to the data.}
    \label{fig:high-model05}
\end{figure*}

\begin{figure*}
	\includegraphics[width=0.8\textwidth]{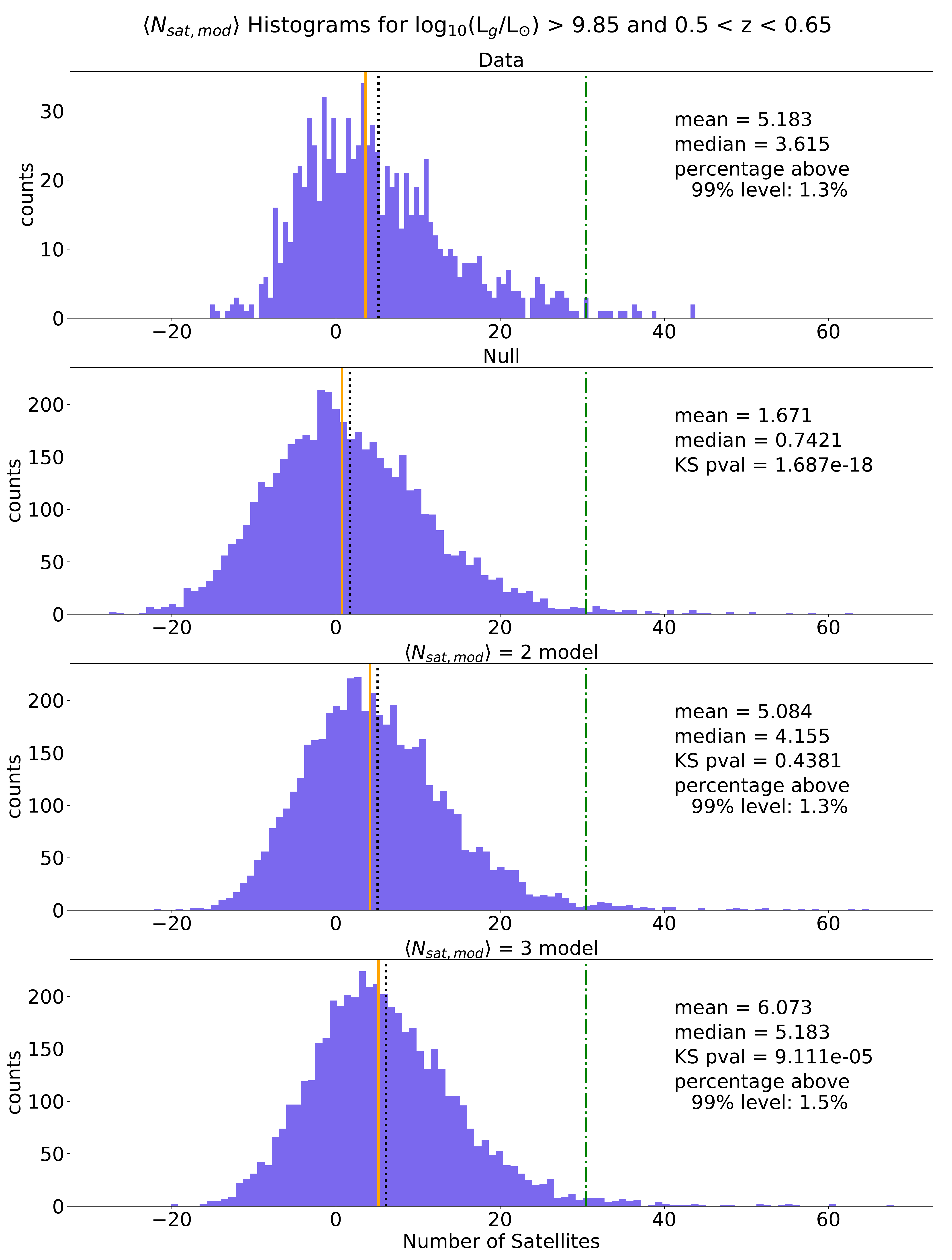}
    \caption{Distributions of the number of satellites found from the data, the null test, and Poisson models for 0.5 $<$ z $<$ 0.65 and log$_{10}$(L) $>$ 9.85. A description of how models are formulated can be found in the text. The dot-dash line indicates the 99 percent confidence. The distribution the data is compared to the null test and the models. The model $\langle N_{sat,mod} \rangle = 2$ is included as an example of a model that is a good fit to the data, according to the 2-sample K-S test, and $\langle N_{sat,mod} \rangle = 3$ is included as an example of model that does not have a good fit to the data.}
    \label{fig:high-model65}
\end{figure*}

Here we show the results of our LRG satellite models based on a Poisson distribution in Figs. \ref{fig:low-model35} through \ref{fig:high-model65}. Models that fit the data with a p-value of 0.05 are shown, as well as examples of models that were not good fits.

\bsp	
\label{lastpage}
\end{document}